\documentclass[12pt]{article}
\usepackage[english]{babel}
\usepackage{graphicx}
\usepackage{amsmath}
\usepackage{amssymb}
\DeclareMathAlphabet{\mathrit}{OT1}{cmr}{m}{sl} 
\DeclareMathAlphabet{\mathritb}{OT1}{cmr}{bx}{sl} 
\usepackage{setspace}
\usepackage{lipsum}
\usepackage{lineno}
\usepackage{parskip}
\usepackage{csquotes}
\usepackage[backend=biber,style=apa,uniquename=false,natbib=true]{biblatex}
\bibliography{references.bib}
\usepackage{parskip}
\usepackage{soul} 
\usepackage{xcolor}
\usepackage{breakcites}
\usepackage[colorlinks=true,linkcolor=blue,citecolor=blue,urlcolor=blue,filecolor=blue]{hyperref}
\usepackage{float}
\usepackage{dutchcal}
\usepackage{subcaption}
\usepackage{upgreek}

\newcommand{\beginsupplement}{
	\setcounter{figure}{0}\renewcommand{\thefigure}{S\arabic{figure}}
	\setcounter{table}{0}\renewcommand{\thetable}{S\arabic{table}}
	\setcounter{equation}{0}\renewcommand{\theequation}{S\arabic{equation}}
}

\usepackage[a4paper, total={6in, 10in}]{geometry}
\title{Conditional multivariate functional PCA for the reconstruction of temperature and salinity profiles partially sampled by deep-diving marine mammals}

\author{Nadège Fonvieille$^{1*}$, Christophe Guinet$^2$, Baptiste Picard$^2$, David Nerini$^1$}
\date{}

\begin{document}

\maketitle
\vspace{-0.5cm}

1. Aix Marseille Univ, Université de Toulon, CNRS, IRD, MIO, Marseille, France
\vspace{-0.3cm}

2. Centre d’Etudes Biologiques de Chizé, Centre National de la Recherche scientifique, 79360 Villiers-en-Bois, France
\vspace{-0.3cm}

*. Corresponding author: Nadège Fonvieille, nadegefonvieille@gmail.com

\subsection*{Orcid}

Nadège Fonvieille: \url{https://orcid.org/0000-0002-2659-4738}
\vspace{-0.3cm}

Christophe Guinet: \url{https://orcid.org/0000-0003-2481-6947}
\vspace{-0.3cm}

Baptiste Picard: \url{https://orcid.org/0000-0001-9565-9446}
\vspace{-0.3cm}

David Nerini: \url{https://orcid.org/0000-0002-7995-1364}

\subsection*{Conflict of interest statement}

The authors declare no conflicts of interest.

\subsection*{Running head}
fPCA to reconstruct truncated TS profiles

\linespread{1.5}\selectfont
\newpage
\begin{abstract}

    We present a statistical method to reconstruct the vertical thermohaline conditions in the Indian Sector of the Southern Ocean, where temperature and salinity profiles are partially sampled by female southern elephant seals. Datasets collected by biologgers provide unprecedented spatial and temporal coverage of ocean conditions. However, the maximum recorded depth varies with the animals' behaviour, offering only a partial view of the vertical environment. Using multivariate functional Principal Component Analysis (PCA), a parametric estimation of the covariance structure and mean function from a set of complete bivariate profiles allows the construction of an eigenfunction basis. By accounting for measurement error variance, partially sampled temperature and salinity profiles can be projected into the eigenspace of the complete profiles through conditional estimation of their functional principal coordinates and then reconstructed over the defined domain. For simulated snippet profiles truncated at depth $z_{\max} = 250$ m and reconstructed over $\mathcal{Z} = [20,500]$ m, reconstruction accuracy increases by 30 \% for temperature and 33 \% for salinity when incorporating geographical covariates. We then reconstruct ~90,000 incomplete profiles from the multivariate functional PCA of ~10,000 profiles reaching 500 m, covering approximately 3 million km$^2$ around the French subantarctic islands.
  
\end{abstract}

Keywords: conditional functional PCA, multivariate functional PCA, partially-sampled curves, oceanography, snippets, thermohaline profiles.

\section{Introduction}

Deep-diving predators equipped with oceanographic sensors allow for ocean sampling at great depths and in remote regions \citep{harcourt2019animal,mcmahon2021animal}.
This approach, known as biologging, is one of the main sources of data in the Southern Ocean \citep{fedak2013impact}, a region particularly difficult to sample with conventional means (e.g., oceanographic vessels and Argo floats; \citealp{riser2016fifteen,johnson2022argo}).
Created in the early 2000s, the MEOP consortium (Marine Mammals Exploring the Oceans Pole to Pole) has gathered over 800,000 temperature and salinity profiles, primarily from the Southern Hemisphere and recorded by pinnipeds during their dives \citep{roquet2014southern}.
Southern elephant seals (\textit{Mirounga leonina} Linnaeus, 1758) represent one of the major contributors to the MEOP database, especially south of 60°S \citep{roquet2013estimates, roquet2017ocean}. When data are recorded at high frequency, tagged animals provide between 60 and 70 profiles per day, reaching depths of up to 2000 m and covering thousands of kilometres over several consecutive months \citep{hindell2016circumpolar}.

As commonly found in oceanographic data, temperature and salinity measurements collected by animal-borne sensors are recorded along a continuum (in this case, depth). Temperature and salinity profiles obtained from the biologgers arrive as sampled curves and fall within the category of functional data. 
Functional Data Analysis (FDA) encompasses a wide range of methods designed for the study of such data \citep{ramsay2005principal}. Over the past two decades, this statistical framework has been increasingly applied in oceanographic research \citep{ariza2022global,ariza2023acoustic,bayle2015moving,assunccao20203d,bourreau2023first,izard2024decomposing,izard2025large,nerini2010cokriging}.
Within FDA, functional Principal Component Analysis (fPCA) is a widely used dimension-reduction technique for projecting curves into a reduced finite-dimensional space, decomposing the signal into modes of variability around a mean trend function. This approach has shown promising results in the description of vertical thermohaline conditions and variability on both a regional and global scale \citep{pauthenet2017linear,pauthenet2019thermohaline,kolbe2021impact}. It has contributed to a better understanding of the variability of ocean fronts \citep{pauthenet2018seasonal}, the distribution of water mass assemblages \citep{fonvieille2023swimming}, and their role in the structure of pelagic communities \citep{tournier2021novel,chevallay2024spies}.

\begin{figure}
	\centering
	\includegraphics[width=.7\linewidth]{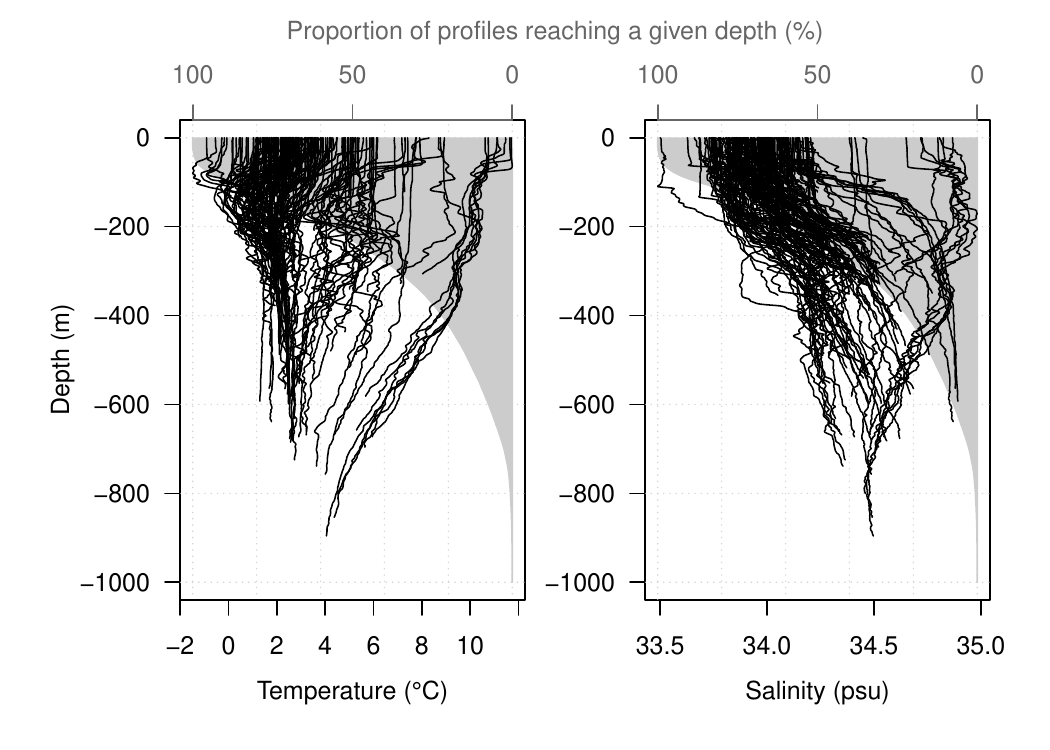}
	\caption{A sample of temperature and salinity profiles recorded by female southern elephant seals of the Kerguelen Archipelago during their post-breeding trips (October to January). The grey polygon in the background indicates the proportion of pelagic profiles (off the plateau) that reach a given depth.}
	\label{FIG:random_profiles}
\end{figure}

The estimation of mean and covariance functions is a fundamental step in FDA, particularly with fPCA \citep{wang2016functional}. This topic has received considerable attention in the literature (e.g., \citealp{lee2002estimating,cai2011optimal,xiao2016fast,zhang2016sparse}). However, smooth estimation remains challenging when measurements are irregular and sparse \citep{yao2005functional,xiao2018fast}, or when each function is observed over a short sub-interval of the entire study domain $\mathcal{Z} = [z_1, z_L] \subset {\rm \mathbb{R}}$ \citep{kneip2020optimal,delaigle2021estimating,lin2021basis,lin2022mean}. 
The latter situation, often referred to as partially-observed data or snippet functions, is intrinsic to the profiles recorded by deep-diving mammals, since the depth reached varies between dives, depending on biophysical features that structure prey distribution \citep{mcintyre2010lifetime}. 
\autoref{FIG:random_profiles} provides an example of such data, displaying some temperature and salinity (TS) profiles sampled along southern elephant seal trajectories, superimposed on the distribution of the maximum depth of dives.
Each of the profiles $X_n$ is observed on a fine grid defined over a dive-dependent depth interval $\mathcal{Z}_n = [z_1, z_{L_n}] \subset \mathcal{Z}$, where $z_1$ represents the sea surface, considered as a fixed lower bound, and $z_{L_n} < z_L$ represents the maximum depth reached.
Recent works have circumvented this constraint by selecting the pool of profiles that reach a desired depth $z_1 < z_{\text{max}} < z_L$ and decomposing them into a suitable basis of functions ranging over the sub-interval $\mathcal{Z}_{max} = [z_1, z_{\text{max}}]$ \citep{tournier2021novel,fonvieille2023swimming,chevallay2024spies,molinet2025linking}. However, the choice of $z_{\text{max}}$ may result in the exclusion of a large number of profiles that do not reach this depth, potentially leading to ecological misinterpretations, as discussed in \citet{fonvieille2023swimming}. 
To overcome this issue, we propose to reconstruct short snippet bivariate TS functions in three steps. The first step uses a dense TS dataset ranging over the interval $\mathcal{Z}$ for the parametric estimation of the mean and covariance functions. An eigenbasis is then constructed from the spectral decomposition of the covariance function. The second step shows how extra partially-observed TS profiles may be projected onto the previous eigenbasis conditionally to random design grid points \citep{yao2005functional}. The third step enhances the reconstruction of partially-observed curves when including some covariates in the fPCA \citep{cardot2007conditional}.

The document is structured as follows. Section 2 provides some technical elements of fPCA and curve projection in the univariate case, which form the basics of our work. Section 3 extends fPCA to bivariate functional data and details the reconstruction of a bivariate TS profile from the projection of its partially-observed data into the eigenspace. The last part of this section incorporates covariates into the projection process to improve the profile reconstruction. In Section 4, we first present a simulation study based on observed bivariate profiles that have been deliberately truncated. We then illustrate the application of the method to an actual dataset recorded by southern elephant seals. Finally, a discussion and future perspectives are presented in Section 5. 

\section{Functional Principal Component Analysis in a nutshell}

\subsection{Karhunen-Loève expansion}

Consider a sample of curves $\mathcal{S}_N = \{X_1, \dots, X_N\}$ that are independent realisations of a smooth random function $\{X(z), z \in \mathcal{Z}\}$, where $\mathcal{Z} = [z_1, z_L]$ is a bounded interval of $\mathbb{R}$. This function belongs to a Hilbert space $\mathcal{X}$ endowed with an inner product $\langle \cdot, \cdot \rangle$ and the associated norm $\| \cdot \|$. The mean function is given with $\mu(z) = \mathrm{E}(X(z))$ and the covariance function is defined as $\gamma(s,z) = \mathrm{Cov}(X(s), X(z)), s, z \in \mathcal{Z}$. Following Mercer’s theorem \citep{hsing2015}, the spectral decomposition of the covariance function provides an orthogonal expansion of $\gamma$ in the form $\gamma(s,z) = \sum_{q\ge 1} \lambda_q\xi_q(s)\xi_q(z), s, z \in \mathcal{Z}$, where pairs $(\lambda_q,\xi_q)_{q\ge 1}$ are real eigenvalues and associated eigenfunctions of the decomposition. The expression of a random curve $X_n \in \mathcal{S}_N$ projected onto that basis is in the form $X_n(z) = \mu(z) + \sum_{q \ge 1} c_{n,q}\xi_q(z)$, where the random coordinates $c_{n,q} = \langle X_n - \mu, \xi_q \rangle, q = 1, 2, \dots$, are uncorrelated random variables known as functional principal components \citep{dauxois1982asymptotic,ramsay2005principal}. They satisfy $\mathrm{E}(c_{n,q}) = 0$ and $\mathrm {Var}(c_{n,q}) = \lambda_q$. If the sequence of positive ordered eigenvalues exhibits rapid decay, it is expected that a good approximation $\widetilde{X}_n^Q$ of $X_n$ may be constructed in a reduced $Q$-dimensional space involving a small number $Q$ of basis functions. The function $\widetilde{X}_n^Q$ is then expressed as a linear combination of the first $Q$ eigenfunctions which form an orthonormal basis in $\mathcal{X}$:
\begin{equation}
	\label{EQ:KLexpansion}
	\widetilde{X}_n^Q(z) = \mu(z) + \sum_{q = 1}^{Q} c_{n,q} \xi_q(z), \quad z \in \mathcal{Z}.
\end{equation}
This decomposition is known as the Karhunen-Loève expansion of $X_n$ truncated at order $Q$ \citep{hsing2015,karhunen1946spektraltheorie,loeve1946fonctions}, also referred to as functional Principal Component Analysis (fPCA). Each curve $X_n$ is characterised by a finite sequence of $Q$ real values, the principal component scores $c_{n,1}, \dots, c_{n,Q}$, making fPCA an efficient method to reduce the dimensionality of high-dimensional datasets.

\subsection{Estimation of the eigencomponents}

Starting with the sample $\mathcal{S}_N$, estimators of the mean $\mu$ and the covariance $\gamma$ functions are given by their empirical versions
\begin{equation*}
    \widehat{\mu}(z) = \frac{1}{N} \sum^N_{n=1} X_n(z), \quad z \in \mathcal{Z},
\end{equation*}
and 
\begin{equation*}
    \widehat{\gamma}(s,z) = \frac{1}{N} \sum^N_{n=1} (X_n(s) - \widehat{\mu}(s))(X_n(z) - \widehat{\mu}(z)), \quad s, z \in \mathcal{Z}.
\end{equation*}

Eigenvalues and corresponding eigenfunctions are obtained by solving the eigenvalue problem
\begin{equation}
	\label{EQ:VP}
    \int_\mathcal{Z} \widehat{\gamma}(s,z) \widehat{\xi}_q(s) \mathrm{d}s = \widehat{\lambda}_q \widehat{\xi}_q(z),
\end{equation}
where $\widehat{\lambda}_1 \geq \dots \geq \widehat{\lambda}_q \geq \dots \geq 0$ are positive eigenvalues, and $(\widehat{\xi}_q)_{q\geq 1}$ are the associated orthonormal eigenfunctions resulting from the spectral decomposition of the empirical covariance function \citep{hsing2015,besse86,ramsay2005principal}. Associated principal component scores can then be estimated from
\begin{equation*}
    \widehat{c}_{n,q} = \int_\mathcal{Z} (X_n(z) - \widehat{\mu}(z)) \widehat{\xi}_q(z) \mathrm{d}z.
\end{equation*}
This estimator allows projecting any extra curve $X_{m}, m > N$ onto the space spanned by the eigenfunctions with
\begin{equation*}
    \widehat{c}_{m,q} = \int_\mathcal{Z} (X_{m}(z) - \widehat{\mu}(z)) \widehat{\xi}_q(z) \mathrm{d}z.
\end{equation*}
A fast decrease in eigenvalues allows the function $X_{m}$ to be well approximated using the finite dimensional basis composed of the first $Q$ eigenfunctions such that
\begin{equation*}
	\widetilde{X}_{m}^Q(z) = \widehat{\mu}(z) + \sum_{q = 1}^{Q} \widehat{c}_{m,q} \widehat{\xi}_q(z), \quad z \in \mathcal{Z}.
\end{equation*}

\subsection{Dealing with a dense dataset}

In practice, curves $X_1, \dots, X_N$ are never recorded as complete curves but rather as discretised measurements. We suppose that observations share the same sampled points $z_{1} < \dots < z_L$ which form a dense design grid on the interval $\mathcal{Z} = [z_1, z_L]$. We further assume that they arrive as noisy observations that can be decomposed such that
\begin{equation*}
	Y_{n,l} = X_{n}(z_l) + \varepsilon_{n,l}, \quad n = 1, \dots, N, \quad l = 1, \dots, L,
\end{equation*}
where $X_n$ is a smooth function and the random error $\varepsilon_{n,l}$ is considered as a measurement error with $\mathrm{E}(\varepsilon_{n,l}) = 0$ and $\mathrm{Var}(\varepsilon_{n,l}) = \sigma^2_{l}$. It is assuming that the smooth functions $X_{n}$ may be expressed as a linear combination of $K$ known basis functions $\phi_1,\dots,\phi_K$ such that
\begin{equation*}
    X_n(z) = \sum_{k=1}^K\alpha_{n,k}\phi_{k}(z), \quad z \in \mathcal{Z}.
\end{equation*}
In this paper, a B-spline basis expansion is adopted \citep{cardot2000nonparametric,Wahba90splines}, an approach widely used in FDA for the flexibility and the versatility of the B-splines \citep{ramsay2005principal,ullah2013applications}. Due to the dense design grid, the coefficients $\alpha_{n,k}, k = 1, \dots, K$, are easily estimated using penalised regression to control the smoothness of the curve \citep{Wahba90splines}. In that case, the curve $X_n$ is approximated with
\begin{equation*}
    \widehat{X}_n(z) = \boldsymbol{\phi}^\prime(z)\widehat{\boldsymbol{\alpha}}_n, \quad n = 1, \dots, N, \quad z \in \mathcal{Z},
\end{equation*}
where $\widehat{\boldsymbol{\alpha}}_n = (\widehat{\alpha}_{n,1}, \dots, \widehat{\alpha}_{n,K})^\prime$ is the vector of estimated coefficients and $\boldsymbol{\phi}(z) = (\phi_1(z), \dots, \phi_K(z))^\prime$ is the vector of the B-spline basis functions evaluated at $z$.

Although measurement errors are often considered as homoscedastic, here we also assume the existence of a smooth variance function $\sigma^2(z)$ such that $\sigma^2_{l} = \sigma^2(z_{l})$. The estimated curves $\widehat{X}_1, \dots, \widehat{X}_N$ which now form the sample at hand, allow for a straightforward estimation of the variance $\sigma^2_l$ over the grid design points $z_1<\dots<z_L$ with
\begin{equation}
    \label{EQ:SIG2_EST}
    \widehat{\sigma}^2(z_l) = \frac{1}{N} \sum^N_{n = 1} (Y_{n,l} - \boldsymbol{\phi}^\prime(z_l)\widehat{\boldsymbol{\alpha}}_n)^2.
\end{equation}

When functions are decomposed into the B-spline basis, estimates of mean and covariance functions turn into 
\begin{equation*}
    \widehat{\mu}(z) = \boldsymbol{\phi}^\prime(z)\overline{\boldsymbol{\alpha}} ~\text{ and }~ \widehat{\gamma}(s,z) = \boldsymbol{\phi}^\prime(s)\mathbf{V}\boldsymbol{\phi}(z),
\end{equation*}
where $\overline{\boldsymbol{\alpha}} = \frac{1}{N}\sum_n\widehat{\boldsymbol{\alpha}}_n$ is the empirical mean of coefficient vectors and $\mathbf{V} = \frac{1}{N}\mathbf{C}^\prime\mathbf{C}$ is the empirical covariance $K \times K$ matrix of the centred coefficients stored in the $N \times K$ matrix $\mathbf{C}$.
Solving the eigenvalue problem in equation (\ref{EQ:VP}) boils down to solving the matrix equation
\begin{equation*}
    \mathbf{W}^{\frac{1}{2}}\mathbf{V}\mathbf{W}^{\frac{1}{2}}\mathbf{b}_k = \widehat{\lambda}_k\mathbf{b}_k,
\end{equation*}
where $\mathbf{W} = \int_\mathcal{Z}\boldsymbol{\phi}(z)\boldsymbol{\phi}^\prime(z) \mathrm{d} z$ is the symetric definite positive Gram matrix of basis functions, $\mathbf{W}^{\frac{1}{2}}$ its square root matrix and $\mathbf{b}_k$ is the eigenvector associated with eigenvalue $\widehat{\lambda}_k$. Eigenfunctions are computed with $\widehat{\xi}_k(z) = \boldsymbol{\phi}^\prime(z)\mathbf{W}^{-1/2}\mathbf{b}_k$. The principal component scores of a function $\widehat{X}_n$ are determined from the elements of the eigen decomposition with
\begin{equation}
    \label{EQ:PCscores_with_coeff}
    \widehat{c}_{n,k} = (\widehat{\boldsymbol{\alpha}}_n - \overline{\boldsymbol{\alpha}})^\prime\mathbf{W}^{\frac{1}{2}}\mathbf{b}_k, \quad k = 1, \dots, K.
\end{equation}

An approximation of that function using the first $Q \le K$ principal component scores is given by
\begin{equation*}
    \widetilde{X}^Q_n = \boldsymbol{\phi}^\prime(z)\left[\overline{\boldsymbol{\alpha}} + \sum_{q=1}^Q\widehat{c}_{n,q}\mathbf{W}^{-\frac{1}{2}}\mathbf{b}_q\right] = \boldsymbol{\phi}^\prime(z)\widetilde{\boldsymbol{\alpha}}^Q_n.
\end{equation*}

Notice that any extra curve  not used to estimate the fPCA eigenelements can also be projected into the eigenspace as an additional observation.

In the following, the \textit{hat} notation will be omitted for the coefficients and the sample $\mathcal{S}_N=\{X_1, \dots, X_N\}$ will correspond to estimated curves with B-splines.

\subsection{Dealing with partially-observed data}

Suppose that we observe an extra curve $X_m\notin\mathcal{S}_N$, which arrives as a set of $L_m$ real pairs $(Y_{m,1}, z_{m,1}), \dots,(Y_{m,L_m}, z_{m,L_m})$. The design points $z_{m,1} < \dots < z_{m,L_m}$ are picked at random from the interval $[z_1, z_L]$ or sampled along a subinterval of $[z_1, z_L]$. 
In such cases, the random curve $X_m$ cannot be projected onto the B-spline basis and therefore, its principal component scores cannot be reliably approximated using equation (\ref{EQ:PCscores_with_coeff}). However, it is possible to estimate the principal component scores $c_{m,q},\dots,c_{m,Q}$ conditionally on the partially-observed values $Y_{m,1},\dots,Y_{m,L_m}$ with
\begin{equation}
    \label{EQ:Yao}
	\widetilde{c}_{m,q} = \widehat{\mathrm{E}}[c_{m,q} | \mathbf{Y}_m] = \widehat{\lambda}_q \widehat{\boldsymbol{\xi}}^\prime_{m,q} \widehat{\boldsymbol{\Sigma}}^{-1}_{m}(\mathbf{Y}_m - \widehat{\boldsymbol{\mu}}_m).
\end{equation}
The vector $\mathbf{Y}_m = (Y_{m,1}, \dots, Y_{m,L_m})^\prime$ contains the noisy observed values of the curve $X_m$, the empirical mean function $\widehat{\boldsymbol{\mu}}_m = (\widehat{\mu}(z_{m,1}), \dots, \widehat{\mu}(z_{m,L_m}))^\prime$ and the estimated $q$-th eigenfunction $\widehat{\boldsymbol{\xi}}_{m,q} = (\widehat{\xi}_q(z_{m,1}), \dots, \widehat{\xi}_q(z_{m,L_m}))^\prime$ are evaluated at design grid points $z_{m,1} < \dots < z_{m,L_m}$. In our case, the entries of the $L_m \times L_m$ covariance matrix $\widehat{\boldsymbol{\Sigma}}_{m}$ are given by 
\begin{equation}
    \label{eq:COVMAT}
    \left(\widehat{\boldsymbol{\Sigma}}_{m}\right)_{j,l} = \widehat\gamma(z_{m,j},z_{m,l}) + \delta_{j,l}\widehat{\sigma}^2(z_{m,l}),
\end{equation}
where $\delta_{j,l} = 1$ if $j = l$, 0 otherwise. Values $\widehat{\sigma}^2(z_{m,l})$ are the variances of the measurement errors given in equation (\ref{EQ:SIG2_EST}). Equation (\ref{EQ:Yao}) was first proposed in \citet{mardia1979multivariate} in a multivariate gaussian framework and was adapted to the functional framework in \citet{yao2005functional}. Compared to the work of \citet{yao2005functional}, we use a parametric estimator of the covariance function $\gamma$ that is based on a B-spline basis decomposition of the complete curves in $\mathcal{S}_N$ before estimating a partially-observed extra curve $X_m\notin\mathcal{S}_N$. Notice that the additional terms that complete the diagonal of the covariance matrix in equation (\ref{eq:COVMAT}) are positive numbers and make this covariance matrix well-conditioned.

Under Gaussian assumptions, the conditional estimator in equation (\ref{EQ:Yao}) provides the best prediction of the principal component scores \citep{mardia1979multivariate}. However, experience shows that the method is still robust when moving away from gaussian assumptions, yielding the best linear predictor \citep{yao2005functional}. 
With $\widehat{\mu}$ and $\widehat{\xi}_{1},\dots,\widehat{\xi}_{K}$ computed from a sufficiently large number $N$ of complete curves $X_1, \dots, X_N$, the partially-observed curve $X_m$ can be approximated with
\begin{equation*}
    \widetilde{X}^Q_m(z)=\widehat{\mu}(z)+\sum_{q=1}^Q\widetilde{c}_{m,q}\widehat{\xi}_q(z),Q<K.
\end{equation*}

\section{Curve projection for bivariate functional data}

We now extend fPCA and the curve projection to partially-observed curves in the case of bivariate functional data. We illustrate the approach using temperature (T ($^\circ\mathrm{C}$)) and salinity (S (psu)) profiles sampled in $\mathcal{Z}$ and decomposed into a B-spline basis.

\subsection{Multivariate fPCA}
\label{SEC:mfPCA}

Consider a sample $\mathcal{S}_N=\{(X_1^T,X_1^S),\dots,(X_N^T,X_N^S)\}$ of $N$ pairs of TS curves  which are functions of depth $z\in\mathcal{Z}=[z_1,z_L]$. Additional covariates, such as location (latitude and longitude) or sampling time, are associated with each pair $(X_n^T,X_n^S)$.

As in the univariate case, these functions are expressed as a linear combination of a finite number $K$ of known B-spline basis functions $\phi_1,\dots,\phi_K$ such that
\begin{equation*}
	X_n^T(z) =  \boldsymbol{\phi}^\prime(z) \boldsymbol{\alpha}^T_n, \quad X_n^S(z) =  \boldsymbol{\phi}^\prime(z) \boldsymbol{\alpha}^S_n, \quad z \in \mathcal{Z},
\end{equation*}
where vectors $\boldsymbol{\alpha}^T_n = (\alpha_{n,1}^T,\dots,\alpha_{n,K}^T)^\prime$ and $\boldsymbol{\alpha}^S_n = (\alpha_{n,1}^S,\dots,\alpha_{n,K}^S)^\prime$ contain the coefficients of the B-spline basis decomposition for temperature and salinity respectively. For the sake of simplicity, the same basis is chosen for temperature and salinity curves but other choices may also be possible.

\begin{figure}[ht]
	\centering
    \includegraphics[width=.7\linewidth]{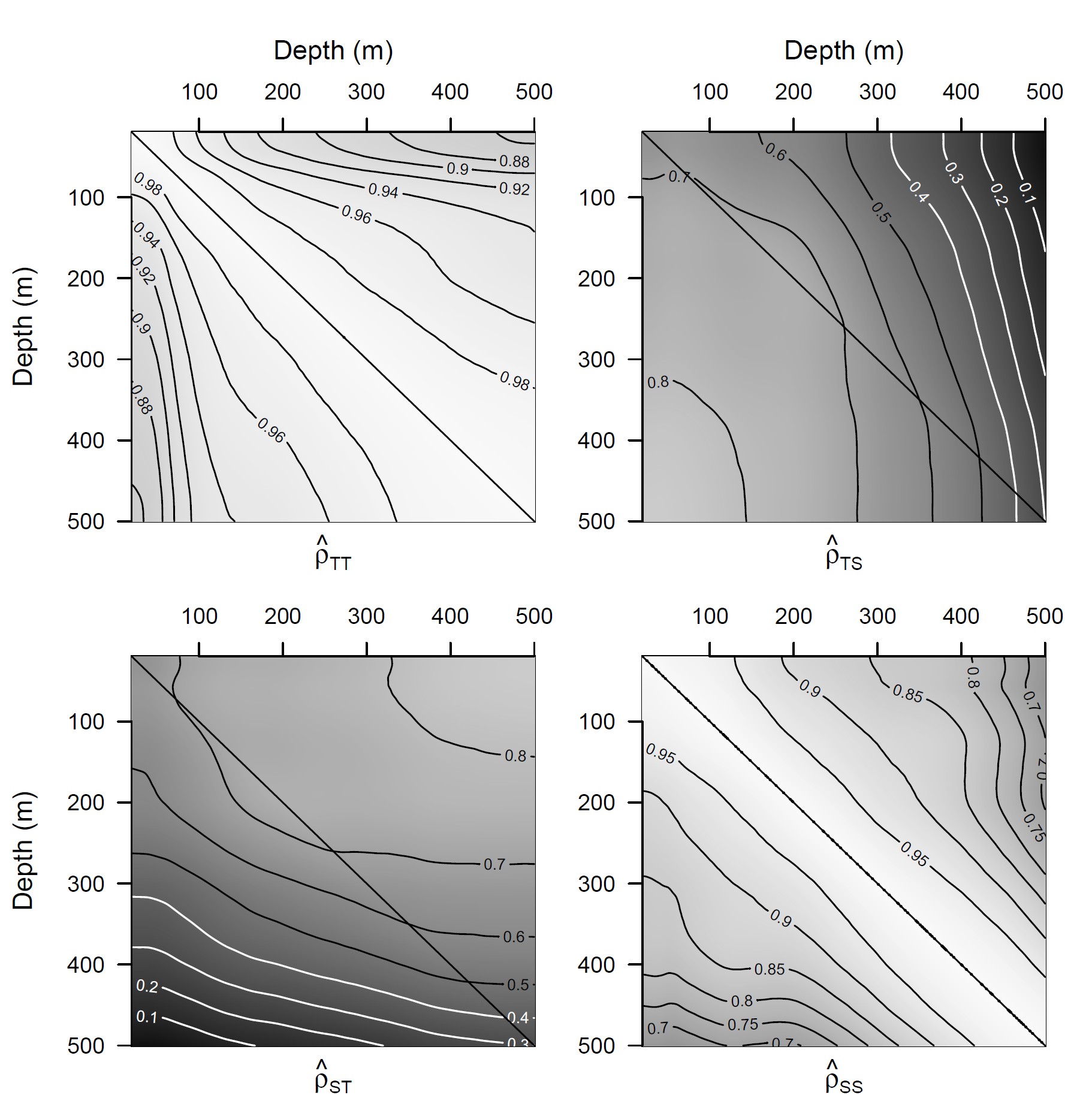}
	\caption{Correlation $(\widehat{\rho}_{TT}, \widehat{\rho}_{SS}$) and cross-correlation $(\widehat{\rho}_{TS},\widehat{\rho}_{ST})$ functions between temperature and salinity estimated for $z \in \mathcal{Z}_{500} = [20, 500]$ m from a sample $\mathcal{S}_N$ of size $N = 10,000$ (see Section \ref{SEC:Data} for data information). Note the symmetry and block structure.}
	\label{FIG:COVAR}
\end{figure}

Consider now the $N\times 2K$ matrix $\mathbf{C}=\left(\mathbf{C}_T,\mathbf{C}_S \right)$ that merges the centred matrices of coefficients for T and S. The $2K\times 2K$ covariance matrix $\mathbf{V}$ of the coefficients is computed with $\mathbf{V} = \frac{1}{N}\mathbf{C}^\prime\mathbf{C}$ and is structured by blocks:
\begin{equation*}
	\mathbf{V} = 
	\begin{bmatrix}
		\mathbf{V}_{TT} & \mathbf{V}_{TS} \\
		\mathbf{V}_{ST} & \mathbf{V}_{SS}
	\end{bmatrix},
\end{equation*}
where $\mathbf{V}_{TS}$ denotes the $K \times K$ cross-covariance matrix between temperature and salinity coefficients used to construct the cross-covariance function $\widehat{\gamma}_{TS}(s,z) = \boldsymbol{\phi}^\prime(s) \mathbf{V}_{TS} \boldsymbol{\phi}(z)$. The \autoref{FIG:COVAR} displays the associated cross-correlation function
\begin{equation*}
    \widehat{\rho}_{TS}(s,z)=\frac{\widehat{\gamma}_{TS}(s,z)}{\sqrt{\boldsymbol{\phi}^\prime(s)\mathbf{V}_{TT}\boldsymbol{\phi}(s)}\sqrt{\boldsymbol{\phi}^\prime(z)\mathbf{V}_{SS}\boldsymbol{\phi}(z)}}    
\end{equation*} 
as well as $\widehat{\rho}_{TT}, \widehat{\rho}_{ST}$ and $\widehat{\rho}_{SS}$, which are easier to interpret than covariance functions.

\begin{figure}[ht]
	\centering
	\includegraphics[width=1\linewidth]{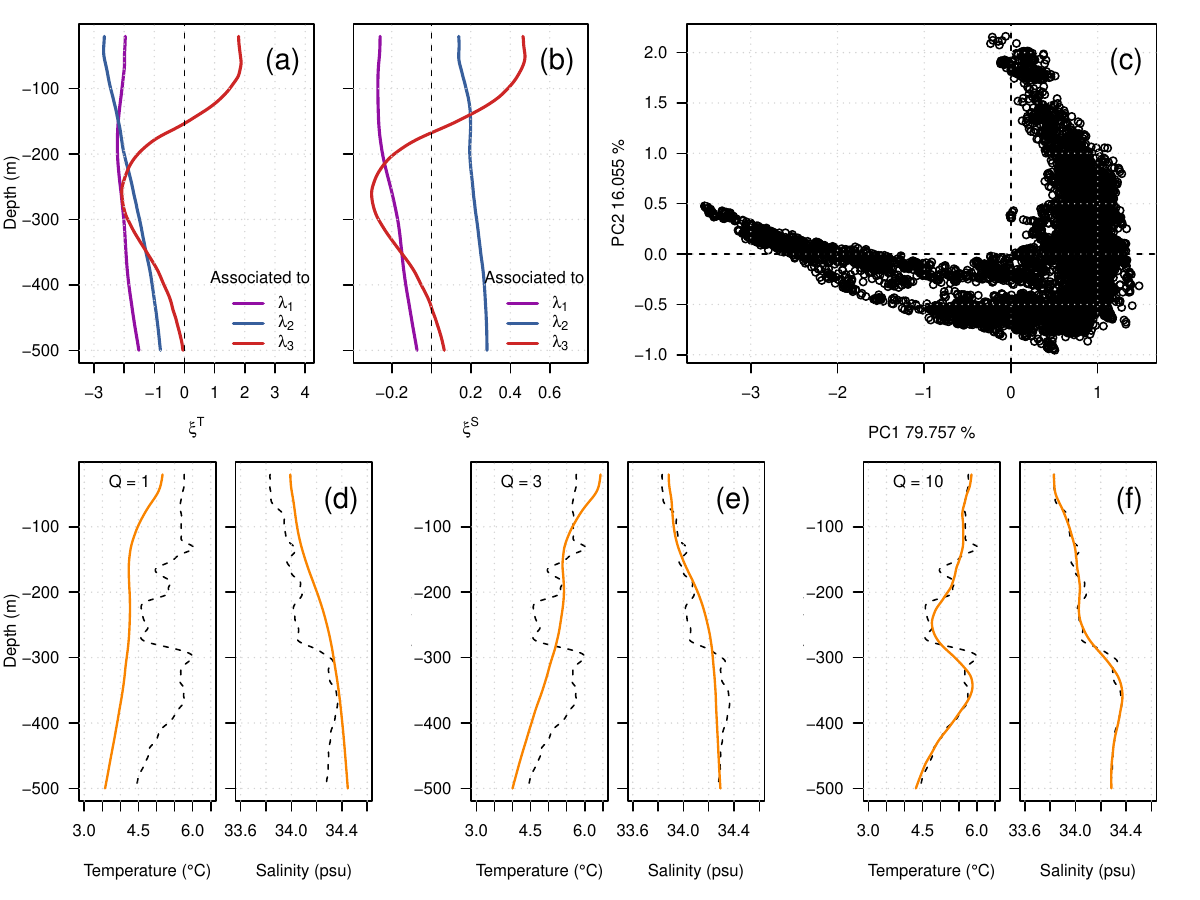}
	\caption{Display of a bivariate functional principal components analysis performed on temperature and salinity functions $\{(X_n^T,X_n^S) \in \mathcal{S}_N, N = 10,000\}$ with depth $z \in [20, 500]$ m (see Section \ref{SEC:Data} for data information). 
	\textbf{(a,b)}. Eigenfunctions $\widehat{\xi}_{1:3}$ associated with the first three eigenvalues $\widehat{\lambda}_{1:3}$. \textbf{(c)}. Principal component scores in the 2D-eigenspace. \textbf{(d,e,f)}. Approximation of TS profiles $(\widetilde{X}_n^T, \widetilde{X}_n^S)$ with $Q \in \{1, 3, 10\}$ (orange lines) and original B-spline profiles $(X^T_n,X^S_n)$ (dotted lines).}
	\label{FIG:xi_PC_Q}
\end{figure}

Dealing with curves projected onto a finite-dimensional B-spline basis, the eigenvalue problem in equation (\ref{EQ:VP}) consists of solving the following eigenequation \citep{nerini2022extending}:
\begin{equation*}
	\mathbf{M}^{1/2} \mathbf{W}^{1/2} \mathbf{V} \mathbf{W}^{1/2} \mathbf{M}^{1/2} \boldsymbol{b}_k = \widehat{\lambda}_k \boldsymbol{b}_k, \quad k = 1, \dots, 2K.
\end{equation*}
The vector $\boldsymbol{b}_k$, of size $2K$, corresponds to the $k$-th eigenvector associated with the eigenvalue $\widehat{\lambda}_k$, and is structured as $\boldsymbol{b}_k = (b^T_{k,1}, \dots, b^T_{k,K}, b^S_{k,1}, \dots, b^S_{k,K})^\prime = (\boldsymbol{b}^{T\prime}_k, \boldsymbol{b}^{S\prime}_k)^\prime$. The $\mathbf{WM}$-orthonormal eigenvector $\boldsymbol{\beta}_k$ can be obtained from $\boldsymbol{b}_k$ such as
$\boldsymbol{\beta}_k = (\boldsymbol{\beta}^{T\prime}_k$, $ \boldsymbol{\beta}^{S\prime}_k)^\prime = \mathbf{M}^{-1/2} \mathbf{W}^{-1/2} \boldsymbol{b}_k$. The eigenvectors are sorted in descending order according to their associated eigenvalue. In the bivariate case, each eigenvector $\boldsymbol{\beta}_k$ generates two eigenfunctions $(\widehat{\xi}_k^T, \widehat{\xi}_k^S)$, also called modes of variability:
\begin{equation*} 
	\left\{ 
	\begin{aligned} 
		\widehat{\xi}^T_k(z) &= \boldsymbol{\phi}^\prime(z) \boldsymbol{\beta}_k^T \\ 
		\widehat{\xi}^S_k(z) &= \boldsymbol{\phi}^\prime(z) \boldsymbol{\beta}_k^S 
	\end{aligned} 
	\right..
\end{equation*} 
The eigenfunctions associated with the largest eigenvalue correspond to the first mode of variability, and so on (e.g., \autoref{FIG:xi_PC_Q}a and b).

The matrix $\mathbf{W}$, of size $2K \times 2K$, is diagonal and also structured by blocks:
\begin{equation*}
	\mathbf{W} = 
	\begin{bmatrix}
		\mathbf{W}_{T} & \boldsymbol{0} \\
		\boldsymbol{0} & \mathbf{W}_{S}
	\end{bmatrix},
\end{equation*}
where the $K \times K$ Gram matrices $\mathbf{W}_T$ and $\mathbf{W}_S$ are symmetric and such that
\begin{equation*}
	\mathbf{W}_T = \mathbf{W}_S = \int_{\mathcal{Z}} \boldsymbol{\phi}(z)\boldsymbol{\phi}^\prime(z) \mathrm{d}z,
\end{equation*}
since the same basis is used to decompose both temperature and salinity profiles. To handle the difference in units between temperature and salinity ($^\circ\mathrm{C}$ and psu), the weighting matrix $\mathbf{M}$, of size $2K \times 2K$, is introduced to normalise the coefficients:
\begin{equation*}
	\mathbf{M} = 
	\begin{bmatrix}
		\mathbf{M}_{T} & \boldsymbol{0} \\
		\boldsymbol{0} & \mathbf{M}_{S}
	\end{bmatrix}.
\end{equation*}
The diagonal entries of matrices $\mathbf{M}_T$ and $\mathbf{M}_S$ are such that
\begin{equation*}
	m_{kk}^T = 1/\tau_T^2, \quad m_{kk}^S = 1/\tau_S^2, \quad k = 1, \dots, K,
\end{equation*}
with $\tau_T^2 = \mathrm{Tr}(\mathbf{V}_{TT}\mathbf{W}_T)$ and $\tau_S^2 = \mathrm{Tr}(\mathbf{V}_{SS}\mathbf{W}_S)$. Matrices $\mathbf{V}_{TT}$ and $\mathbf{V}_{SS}$ correspond to the $K \times K$ covariance matrices of the temperature and salinity coefficients, respectively. 

As in the univariate case, observations can be projected into a low-dimensional space $Q \le 2K$ corresponding to the basis formed by the first $Q$ modes of variability (\autoref{FIG:xi_PC_Q}c). The coordinates of the profiles in the eigenspace are given by the principal component coordinates $\mathbf{c}_{n} = (c_{n,1}, \dots, c_{n,Q})^\prime$ by computing 
\begin{equation*} 
	\mathbf{P}_Q = \mathbf{C} \mathbf{W} \mathbf{M} \mathbf{B}_Q, \label{EQ:PCs} 
\end{equation*} 
where $\mathbf{P}_Q = (\mathbf{c}_{1}^\prime, \dots, \mathbf{c}_{N}^\prime)^\prime$ is the matrix of size $N \times Q$ containing the first $Q$ principal component scores. Matrix $\mathbf{B}_Q$, of size $2K \times Q$, contains the $\mathbf{WM}$-orthonormal eigenvectors structured by blocks:
\begin{equation*} 
	\mathbf{B}_Q = 
	\begin{bmatrix} 
		\mathbf{B}_Q^T \\ 
        \mathbf{B}_Q^S 
	\end{bmatrix},
\end{equation*}
with $\mathbf{B}_Q^T = (\boldsymbol{\beta}^T_1,\dots, \boldsymbol{\beta}^T_Q)$ and $\mathbf{B}_Q^S = (\boldsymbol{\beta}^S_1,\dots, \boldsymbol{\beta}^S_Q)$. Following equation (\ref{EQ:KLexpansion}), each pair $(X_n^T, X_n^S)$ can finally be approximated by summing the first $Q$ eigenfunctions weighted by the corresponding principal coordinates: 
\begin{equation}
	\label{EQ:KLexpansion2}
	\left\{ 
	\begin{aligned} 
		\widetilde{X}_n^T(z) &= \boldsymbol{\phi}^\prime(z)
		[\boldsymbol{\overline{\alpha}}_T + \mathbf{B}_Q^T \mathbf{c}_n] = \boldsymbol{\phi}^\prime(z) \boldsymbol{\widetilde{\alpha}}_n^T \\ 
		\widetilde{X}_n^S(z) &= \boldsymbol{\phi}^\prime(z)
		[\boldsymbol{\overline{\alpha}}_S + \mathbf{B}_Q^S \mathbf{c}_n] = \boldsymbol{\phi}^\prime(z) \boldsymbol{\widetilde{\alpha}}_n^S
	\end{aligned}
	\right.,
\end{equation}

where $\boldsymbol{\overline{\alpha}}_T$ and $\boldsymbol{\overline{\alpha}}_S$ are mean coefficient vectors for T and S. \autoref{FIG:xi_PC_Q} displays the approximation of a bivariate observation $(X_n^T,X_n^S)$ using a varying number $Q$ of eigenfunctions.

\subsection{Conditional estimation for partially-observed curves}

Suppose a new bivariate observation $(X_m^T, X_m^S) \notin \mathcal{S}_N$ that arrives as a set of noisy temperature-salinity tuples $\{(Y^T_{m,l}, Y^S_{m,l}), l = 1, \dots, L_m\}$ measured at random design grid points $z_{m,1} < \dots < z_{m,L_m}$, with $z_{m,L_m} < z_L$. Equation (\ref{EQ:Yao}) can be extended to the bivariate case with
\begin{equation}
    \label{EQ:Yao2D}
    \widetilde{c}_{m,q} = \widehat{\lambda}_q(\widehat{\boldsymbol{\xi}}^{T\prime}_{m,q}, \widehat{\boldsymbol{\xi}}^{S\prime}_{m,q})\widehat{\boldsymbol{\Sigma}}_m^{-1} \left(
        \begin{array}{c}
            \mathbf{Y}_m^T-\widehat{\boldsymbol{\mu}}_m^T\\
            \mathbf{Y}_m^S-\widehat{\boldsymbol{\mu}}_m^S
        \end{array}
    \right).
\end{equation}
Denoting $\boldsymbol{\Phi}_m = (\boldsymbol{\phi}^\prime(z_{m,1}), \dots, \boldsymbol{\phi}^\prime(z_{m,L_m}))^\prime$ the $L_m \times K$ matrix of B-spline basis functions evaluated at design points $z_{m,1} < \dots < z_{m,L_m}$, $\widehat{\boldsymbol{\xi}}^{T}_{m,q} = \boldsymbol{\Phi}_m\boldsymbol{\beta}^T_q$ and $\widehat{\boldsymbol{\xi}}^{S}_{m,q} = \boldsymbol{\Phi}_m\boldsymbol{\beta}^S_q$ are the vectors of the temperature and salinity eigenfunctions evaluated at the design points. The vectors $\mathbf{Y}^{T}_m = (Y^T_{m,1}, \dots, Y^T_{m,L_m})^\prime$ and $\mathbf{Y}^{S}_m = (Y^S_{m,1}, \dots, Y^S_{m,L_m})^\prime$ are the observed values for temperature and salinity, and the vectors $\widehat{\boldsymbol{\mu}}_m^{T} = \boldsymbol{\Phi}_m\overline{\boldsymbol{\alpha}}_T$  and $\widehat{\boldsymbol{\mu}}_m^{S} = \boldsymbol{\Phi}_m\overline{\boldsymbol{\alpha}}_S$ are the mean temperature and salinity curves evaluated at design points. The $2L_m \times 2L_m$ matrix $\widehat{\boldsymbol{\Sigma}}_{m}$ is structured by blocks:
\begin{equation*}
	\widehat{\boldsymbol{\Sigma}}_{m} =
	\begin{bmatrix}
	    \mathbf{G}^m_{TT} & \mathbf{G}^m_{TS} \\
		\mathbf{G}^m_{ST} & \mathbf{G}^m_{SS}
	\end{bmatrix}
	+ \mathrm{diag}\{\widehat{\boldsymbol{\sigma}}_m^2\}.
\end{equation*}
Entries of the matrix $\mathbf{G}^m_{TS}$ of size $L_m \times L_m$ are 
\begin{equation*}
	(\mathbf{G}^m_{TS})_{j,l} = \widehat{\gamma}_{TS}(z_{m,j},z_{m,l}) = \boldsymbol{\phi}^\prime(z_{m,j})\mathbf{V}_{TS}\boldsymbol{\phi}(z_{m,l}), \quad j, l = 1, \dots, L_m,
\end{equation*}
where the $K \times K$ matrix $\mathbf{V}_{TS}$ corresponds to the cross-covariance matrix of the temperature and salinity coefficients estimated from the sample of $N$ complete pairwise curves. The vector $\widehat{\boldsymbol{\sigma}}_m^2 = (\widehat{\sigma}^2_T(z_{m,1}), \dots, \widehat{\sigma}^2_T(z_{m,L_m}), \widehat{\sigma}^2_S(z_{m,1}), \dots, \widehat{\sigma}^2_S(z_{m,L_m}))^\prime$ gathers the variance of the measurement error evaluated at the design points $z_{m,1} < \dots < z_{m,L_m}$ for the temperature and salinity. 

The temperature and salinity curves of the partially-observed subject are predicted over the whole domain $\mathcal{Z}$ when estimating the coefficients:
\begin{equation}
    \label{EQ:KLexpansion_noCV}
	\left\{ 
	\begin{aligned} 
		\widetilde{X}_m^T(z) &= \boldsymbol{\phi}^\prime(z)
		[\boldsymbol{\overline{\alpha}}_T + \mathbf{B}_Q^T \widetilde{\mathbf{c}}_m] = \boldsymbol{\phi}^\prime(z) \boldsymbol{\widetilde{\alpha}}_m^T \\ 
		\widetilde{X}_m^S(z) &= \boldsymbol{\phi}^\prime(z)
		[\boldsymbol{\overline{\alpha}}_S + \mathbf{B}_Q^S \widetilde{\mathbf{c}}_m] = \boldsymbol{\phi}^\prime(z) \boldsymbol{\widetilde{\alpha}}_m^S
	\end{aligned}
	\right..
\end{equation}
Note that coefficients of the mean functions, covariance matrices and eigenelements are estimated using the sample $\mathcal{S}_N$ of complete curves.

\subsection{Adding a covariate effect}

According to the approach proposed in \citet{cardot2007conditional}, it is possible to incorporate the effect of a quantitative covariate to improve the fPCA reconstitution of an observed curve. In our case, an fPCA is carried out on the $N$ fully-observed bivariate profiles conditionally to the spatial location $\boldsymbol{\Upsilon} = (\Upsilon_{\text{lon}}, \Upsilon_{\text{lat}})^\prime$ associated with an extra partially-observed curve.

 Start with a partially-observed curve $(X^T_m,X^S_m), m > N$ located at position $\boldsymbol{\Upsilon} = \boldsymbol{\upsilon}_{m}$. The shape of $(X^T_m,X^S_m)$ is estimated from the eigenelements obtained from the fPCA performed on the sample $\{(X_1^T, X_1^S), \dots, (X_N^T, X_N^S)\}$, where the mean and covariance functions are estimated by modifying the weight of the observations conditionally to the position $\boldsymbol{\Upsilon} = \boldsymbol{\upsilon}_{m}$. The estimator of the mean functions is then defined as
\begin{equation*}
	\widehat{\mu}_T(\boldsymbol{\upsilon}_{m},z) = \boldsymbol{\phi}^\prime(z) \overline{\boldsymbol{\alpha}}_T(\boldsymbol{\upsilon}_{m}), \quad  \widehat{\mu}_S(\boldsymbol{\upsilon}_{m},z) = \boldsymbol{\phi}^\prime(z) \overline{\boldsymbol{\alpha}}_S(\boldsymbol{\upsilon}_{m}),
\end{equation*}
where the conditional empirical mean coefficients are given by  
$\boldsymbol{\overline{\alpha}}_T(\boldsymbol{\upsilon}_{m}) = \sum_{n=1}^{N} \omega_n(\boldsymbol{\upsilon}_{m},h)\boldsymbol{\alpha}_n^T$ for the temperature and $\boldsymbol{\overline{\alpha}}_S(\boldsymbol{\upsilon}_{m}) = \sum_{n=1}^{N} \omega_n(\boldsymbol{\upsilon}_{m},h)\boldsymbol{\alpha}_n^S$ for the salinity. 
The weighted covariance matrix between coefficients becomes $\mathbf{V}_{m} = \mathbf{C}_{m}^{\prime} \boldsymbol{\Omega}_{m} \mathbf{C}_{m}$, with $\mathbf{C}_{m}$ the matrix of coefficients centred around the conditional mean $\boldsymbol{\overline{\alpha}}(\boldsymbol{\upsilon}_{m}) = (\boldsymbol{\overline{\alpha}}_{T}^{\prime}(\boldsymbol{\upsilon}_{m}), \boldsymbol{\overline{\alpha}}_{S}^{\prime}(\boldsymbol{\upsilon}_{m}))^\prime$ and the weighting diagonal matrix $\boldsymbol{\Omega}_m = \mathrm{diag}\{\omega_1(\boldsymbol{\upsilon}_{m},h), \dots, \omega_N(\boldsymbol{\upsilon}_{m},h)\}$. 
Weights $\omega_n$ are such that
\begin{equation*}
	\omega_n(\boldsymbol{\upsilon}_{m},h) = \frac{K_{h}(\boldsymbol{\Upsilon}_n - \boldsymbol{\upsilon}_{m})}{\sum_{j=1}^{N} K_{h}(\boldsymbol{\Upsilon}_j - \boldsymbol{\upsilon}_{m})},
\end{equation*}
where $K_h$ is a kernel function centred at $\boldsymbol{\upsilon}_{m}$ and $h$ is a smoothing bandwidth. Choices of $K_h$ and bandwidth $h$ are further discussed in the following section. 

Eigenfunctions and eigenvalues can be estimated conditionally on spatial position $\boldsymbol{\upsilon}_{m}$, following Section \ref{SEC:mfPCA}, which in turn allows the estimation of the conditional principal coordinates of the partially-observed pairs $(X^T_m,X^S_m)$ with 
\begin{equation*}
    \widetilde{c}_{m,q} = \widehat{\lambda}_q(\boldsymbol{\upsilon}_{m})(\widehat{\boldsymbol{\xi}}^{T\prime}_{m,q}, \widehat{\boldsymbol{\xi}}^{S\prime}_{m,q})(\boldsymbol{\upsilon}_{m})\widehat{\boldsymbol{\Sigma}}_m^{-1}(\boldsymbol{\upsilon}_{m}) \left(
        \begin{array}{c}
            \mathbf{Y}_m^T-\widehat{\boldsymbol{\mu}}_m^T(\boldsymbol{\upsilon}_{m})\\
            \mathbf{Y}_m^S-\widehat{\boldsymbol{\mu}}_m^S(\boldsymbol{\upsilon}_{m})
        \end{array}
    \right).
\end{equation*}
Finally, equation (\ref{EQ:KLexpansion2}) can be adapted to predict the partially-observed bivariate profile $(X_{m}^T,X_{m}^S)$ over $\mathcal{Z}$, conditionally on its location $\boldsymbol{\upsilon}_{m}$:
\begin{equation}
	\label{EQ:KLexpansion3}
	\left\{ 
	\begin{aligned} 
		\widetilde{X}_{m}^T(\boldsymbol{\upsilon}_{m},z) &= \boldsymbol{\phi}^\prime(z) 
		[\boldsymbol{\overline{\alpha}}_T(\boldsymbol{\upsilon}_{m}) + \mathbf{B}_Q^T(\boldsymbol{\upsilon}_{m}) \widetilde{\mathbf{c}}_{m}] = \boldsymbol{\phi}^\prime(z) \boldsymbol{\widetilde{\alpha}}_m^T(\boldsymbol{\upsilon}_{m}) \\ 
		\widetilde{X}_{m}^S(\boldsymbol{\upsilon}_{m},z) &= \boldsymbol{\phi}^\prime(z) 
		[\boldsymbol{\overline{\alpha}}_S(\boldsymbol{\upsilon}_{m}) + \mathbf{B}_Q^S(\boldsymbol{\upsilon}_{m}) \widetilde{\mathbf{c}}_{m}] = \boldsymbol{\phi}^\prime(z) \boldsymbol{\widetilde{\alpha}}_m^S(\boldsymbol{\upsilon}_{m})
	\end{aligned}
	\right.,
\end{equation}
where $\widetilde{\mathbf{c}}_{m} = (\widetilde{c}_{m,1}, \dots, \widetilde{c}_{m,Q})^\prime$ is the vector of principal coordinates of $(X_{m}^T,X_{m}^S)$. 

Consider now 
\begin{equation*}
         \mathrm{JISE}_m = \frac{1}{\tau^2_T}\lVert X^T_m - \widetilde{X}^T_m\rVert^2 + \frac{1}{\tau^2_S}\lVert X^S_m - \widetilde{X}^S_m\rVert^2
\end{equation*}
the joint integrated squared error between observation $(X^T_m, X^S_m)$ and its estimation $(\widetilde{X}^T_m,\widetilde{X}^S_m)$ and
\begin{equation*}
         \mathrm{JISE}_m(\boldsymbol{\upsilon}_{m}) = \frac{1}{\tau^2_T(\boldsymbol{\upsilon}_{m})}\lVert X^T_m - \widetilde{X}^T_m(\boldsymbol{\upsilon}_{m})\rVert^2 + \frac{1}{\tau^2_S(\boldsymbol{\upsilon}_{m})}\lVert X^S_m-\widetilde{X}^S_m(\boldsymbol{\upsilon}_{m})\rVert^2,
\end{equation*}
the joint integrated squared error between $(X^T_m, X^S_m)$ and its estimation $(\widetilde{X}^T_m(\boldsymbol{\upsilon}_{m}),\widetilde{X}^S_m(\boldsymbol{\upsilon}_{m}))$ with covariate. It is expected that, for every dimension $Q \le 2K$,
\begin{equation*}
    \mathrm{E}\left(\mathrm{JISE}_m(\boldsymbol{\upsilon}_{m}) |\boldsymbol{\Upsilon} = \boldsymbol{\upsilon}_{m} \right) \le \mathrm{E}\left(\mathrm{JISE}_m\right)    
\end{equation*}
 as pointed out in \citet{cardot2007conditional}, equation 8.

\section{Application}

The reconstruction method described above is applied to temperature and salinity profiles recorded simultaneously along southern elephant seal dives. This section is divided into two parts. The first focuses on estimating the principal component scores $\widetilde{c}_{m,q}$ conditionally on the observed values of TS profiles, and on evaluating the reconstruction performance for simulated truncated bivariate profiles. The second part presents the reconstruction of profiles collected along a southern elephant seal trajectory with variable dive depths. In both cases, the reconstructions are performed down to 500 m. This depth was chosen based on (i) oceanographic knowledge (eddies in the Southern Ocean extend to at least this depth \citep{siegelman2020enhanced,siegelman2025ubiquity} and (ii) female elephant seal behaviour (the 75th percentile of maximum dive depths in pelagic dives is around 500 m). This choice also ensures that a sufficient number of profiles are retained to estimate the eigenelements.

\subsection{Data}
\label{SEC:Data}

The following two sections rely on a dataset of 111,289 TS profiles collected off the Kerguelen Plateau by 24 female southern elephant seals during their post-breeding trips (October to January) in 2018, 2019, and 2020. TS data were recorded by CTD biologgers (Conductivity–Temperature–Depth), together with time and satellite-based locations (Argos or GPS). Some locations are missing due to transmitter failure or poor location accuracy. Tag deployment and retrieval procedures are detailed in \citet{mcmahon2000field}. All scientific protocols were approved by the Ethics Committee ComEth Anses/ENVA/UPEC (No.~16-078-2016100709156983; 19-040 \#21375) and have been shown not to affect short-term weight gain or long-term survival \citep{mcmahon2008tracking}.
The CTD tags recorded high resolution data at 0.5 Hz (one measurement every 2 s), including pressure ($\pm$ 2 dbar), temperature ($\pm$ 0.005 $^\circ\mathrm{C}$), and conductivity ($\pm$ 0.005 mS cm$^{-1}$) \citep{boehme2009animal}. All raw data were corrected following the procedure described on the MEOP website (\url{https://meop.net}). Following processing and validation, temperature and salinity profiles extend to depths of up to 1000 m, with a vertical resolution of 1 m and final accuracies of $\pm$ 0.02 $^\circ\mathrm{C}$ and $\pm$ 0.03 psu.

\subsection{Simulation Study}

\begin{figure} [H]
	\centering
	\includegraphics[width=1\linewidth]{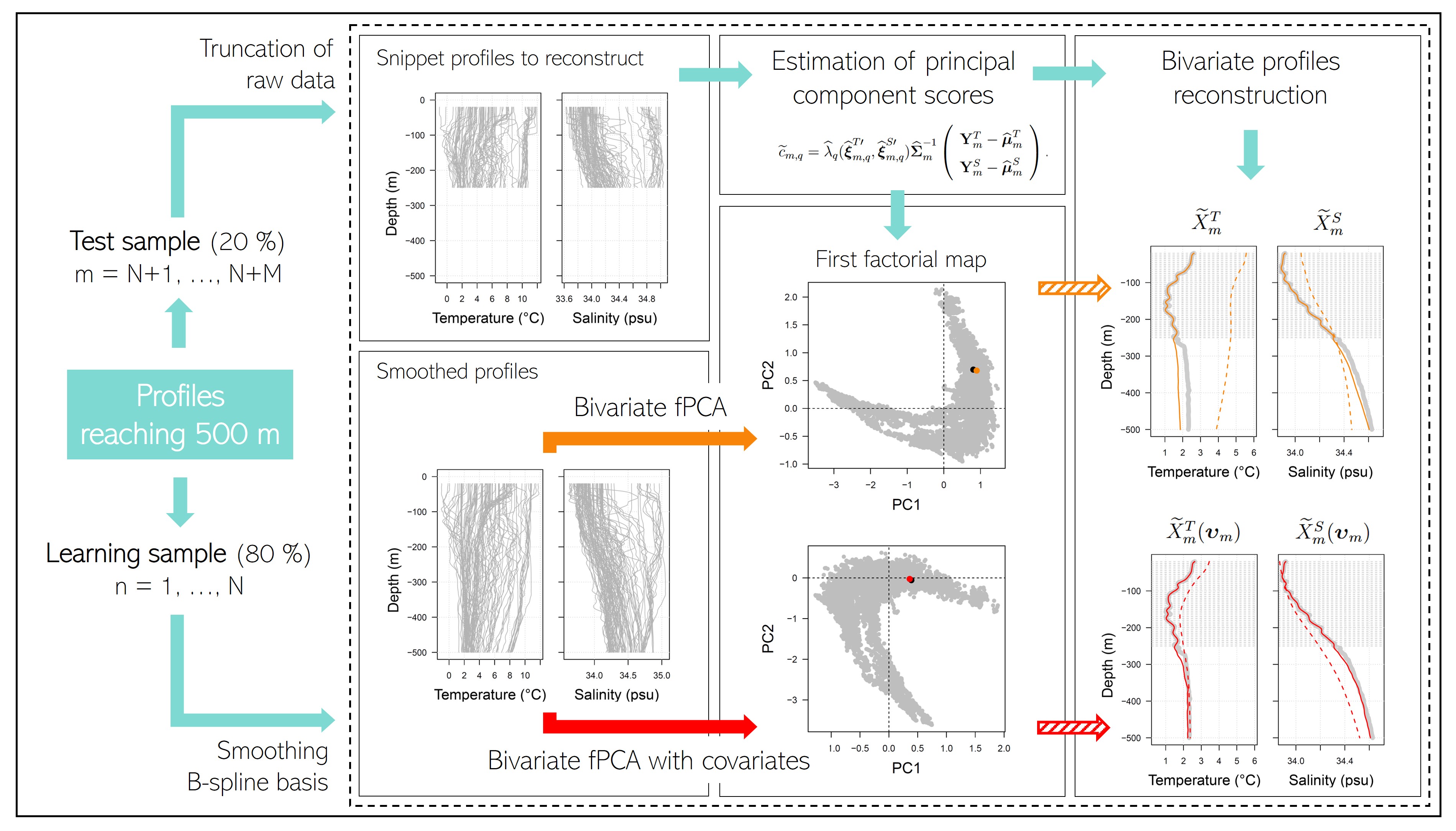}
	\caption{Workflow illustrating the reconstruction steps for simulated degraded bivariate profiles. The dataset is divided into training and test samples. The test sample consists of truncated raw profiles, while the training set is composed with smoothed profiles obtained through B-spline decomposition. Principal component scores $\widetilde{c}_{m,q}$ and profile shapes are then estimated conditionally on the observed values $\{(Y^T_{m,l}, Y^S_{m,l}), l = 1, \dots, L_m\}$, either without (orange path) or with (red path) geographical location included as a covariate in the fPCA. The dashed lines in the right-hand panel represent the mean profiles. Note that the 2D map of the fPCA differs when covariates are included. The estimated principal component scores (red dot) lie closer to the conditional mean (coordinates $(0,0)$ in the factorial map).}
	\label{FIG:WF}
\end{figure}

In this section, we simulate truncated profiles using the dataset described above and evaluate the reconstruction error down to 500 m, according to the snippet depth and the inclusion of covariates in the prediction process. The approach follows the workflow presented in \autoref{FIG:WF}.

\textit{Snippet profiles}

Consider the sample $\{(X^T_1, X^S_1), \dots, (X^T_{N_{500}}, X^S_{N_{500}})\}$ of size $N_{500} = 13,986$, expressed as a linear combination of $K = 50$ B-spline basis functions over the interval $\mathcal{Z}_{500} = [20,500]$ m and spatially located. Select at random a subsample of size $N$ which represents 80 \% of data: it is used to estimate the eigenelements of the multivariate fPCA. The remaining 20 \% of data constitute a test sample of size $M=N_{500}-N$ used to evaluate the performance of the reconstruction method. The raw data of these functions are successively truncated along a sequence of maximum depth $z_{max} \in \{50, \dots, 490\}$ m with $20$ m increments. Each resulting snippet function is thus reconstructed with the available raw data over the interval $\mathcal{Z}_{500} = [20,500]$ m. 

\textit{Covariates effect on the fPCAs}

For each pairwise curves $(X_m^T, X_m^S)$ belonging to the test sample, the reconstruction is performed both conditionally and not-conditionally on location, using the geographical coordinates $\boldsymbol{\upsilon}_{m} = (\text{lon}_m,\text{lat}_m)^\prime$ (red and orange paths in \autoref{FIG:WF}). In the conditional case, a bivariate Gaussian kernel with a diagonal covariance matrix $h^2\mathbf{I}_2$ is used to weight the $N$ profiles according to their geographic proximity to $X_m$. In this study, we assume no correlation between latitude and longitude and the value of $h$ matches the typical size of mesoscale physical structures such as eddies (i.e., 100 km; \citealp{levy2018role}). 
Note that the conditional approach requires the fPCA to be run on the $N$ bivariate curves for every estimation of a bivariate profile $(X^T_m, X^S_m), m \in \{N+1, \dots, N+M\}$ since the weight of an observation is conditional on the covariate value.

\textit{Reconstruction errors}

The joint integrated squared error $\mathrm{JISE}_m$ between the observation $(X^T_{m}, X^S_{m})$ and its prediction $(\widetilde{X}^T_{m}, \widetilde{X}^S_{m})$ is computed in both cases (with or without covariate conditioning) with
\begin{equation}
	\label{EQ:dist}
	\mathrm{JISE}_m = \frac{1}{\tau_T^2} (\boldsymbol{\alpha}_{m}^T - \widetilde{\boldsymbol{\alpha}}_{m}^T)^\prime \mathbf{W}_T (\boldsymbol{\alpha}_{m}^T - \widetilde{\boldsymbol{\alpha}}_{m}^T) + \frac{1}{\tau_S^2} (\boldsymbol{\alpha}_{m}^S - \widetilde{\boldsymbol{\alpha}}_{m}^S)^\prime \mathbf{W}_S (\boldsymbol{\alpha}_{m}^S - \widetilde{\boldsymbol{\alpha}}_{m}^S),
\end{equation} 
where $\boldsymbol{\alpha}_m^T$ and $\boldsymbol{\alpha}_m^S$ are the coefficient vectors of the B-spline decomposition for T and S, and $\widetilde{\boldsymbol{\alpha}}_{m}^T$ and $\widetilde{\boldsymbol{\alpha}}_{m}^S$ are the coefficient TS vectors approximated using the principal component scores $\widetilde{c}_{m,1}, \dots, \widetilde{c}_{m,Q}$ (see equation (\ref{EQ:KLexpansion_noCV}) or (\ref{EQ:KLexpansion3})).

The $\mathrm{JISE}_m$ can be split into two interpretable error measures for temperature and salinity respectively with
\begin{equation}
    \begin{aligned} 
    	\mathrm{ISE}_m^T &= \frac{1}{z_L-z_1} \left[(\boldsymbol{\alpha}_{m}^T - \widetilde{\boldsymbol{\alpha}}_{m}^T)^\prime \mathbf{W}_T (\boldsymbol{\alpha}_{m}^T - \widetilde{\boldsymbol{\alpha}}_{m}^T)\right]\\
        \mathrm{ISE}_m^S &= \frac{1}{z_L-z_1} \left[(\boldsymbol{\alpha}_{m}^S - \widetilde{\boldsymbol{\alpha}}_{m}^S)^\prime \mathbf{W}_S (\boldsymbol{\alpha}_{m}^S - \widetilde{\boldsymbol{\alpha}}_{m}^S)\right],
    \end{aligned}
\end{equation}
where $z_1$ and $z_L$ denote the lower and upper bounds of the depth interval. We report $\mathrm{IE}_m^T = \sqrt{\mathrm{ISE}_m^T}$ and
$\mathrm{IE}_m^S = \sqrt{\mathrm{ISE}_m^S}$, which have the same units as T ($^\circ\mathrm{C}$) and S (psu) respectively.

\textit{Confidence intervals}

Asymptotic pointwise confidence intervals are constructed for $(\widetilde{X}_{m}^T, \widetilde{X}_{m}^S)$, following \citet{yao2005functional} and assuming that both functions may be well approximated by the first $Q$ eigenfunctions:
\begin{equation*}
    \left\{
    \begin{array}{c}
        \widetilde{X}_{m}^T(z) \pm \upvarphi^{-1}(1-\alpha/2) \sqrt{\widehat{\boldsymbol{\xi}}_Q^{T\prime}(z) \mathbf{Q}_T \widehat{\boldsymbol{\xi}}_Q^T(z)}\\
        \widetilde{X}_{m}^S(z) \pm \upvarphi^{-1}(1-\alpha/2) \sqrt{\widehat{\boldsymbol{\xi}}_Q^{S\prime}(z) \mathbf{Q}_S \widehat{\boldsymbol{\xi}}_Q^S(z)}
    \end{array}
	\right.,
\end{equation*}
with $\upvarphi$ the standard Gaussian cumulative distribution function, $\widehat{\boldsymbol{\xi}}_Q^T(z) = (\widehat{\xi}^T_1(z), \dots$, $\widehat{\xi}^T_Q(z))^\prime$ and $\widehat{\boldsymbol{\xi}}_Q^S(z) = (\widehat{\xi}^S_1(z), \dots, \widehat{\xi}^S_Q(z))^\prime$ the vectors of T and S eigenfunctions truncated at order $Q$ and evaluated at $z$. The matrix $\mathbf{Q}_T$ is computed as $\mathbf{Q}_T = \widehat{\mathbf{\Lambda}} - \widehat{\mathbf{H}}_T\widehat{\boldsymbol{\Sigma}}^{-1}_{m} \widehat{\mathbf{H}}_T^\prime$, where $\widehat{\mathbf{\Lambda}} = \text{diag}\{\widehat{\lambda}_1, \dots, \widehat{\lambda}_Q\}, \widehat{\mathbf{H}}_T = (\widehat{\lambda}_1\widehat{\boldsymbol{\xi}}^T_{m,1}, \dots, \widehat{\lambda}_Q\widehat{\boldsymbol{\xi}}^T_{m,Q})^\prime$ and $\widehat{\boldsymbol{\xi}}^T_{m,q} = (\widehat{\xi}^T_q(z_{m,1})$, $\dots, \widehat{\xi}^T_q(z_{m,L_{m}}))^\prime$. Entries of the covariance matrix $\widehat{\boldsymbol{\Sigma}}_m$ are 
\begin{equation*}
    \left(\widehat{\boldsymbol{\Sigma}}_{m}\right)_{j,l} = \widehat\gamma_{TT}(z_{m,j},z_{m,l})+\delta_{j,l}\widehat{\sigma}_T^2(z_{m,l}),
\end{equation*}
where $\delta_{j,l} = 1$ if $j = l$ and 0 otherwise. Matrix $\mathbf{Q}_S$ is obtained by replacing T by S in the above expressions.

\textit{Overall results}

\begin{figure}[H]
	\centering
	\includegraphics[width=1\linewidth]{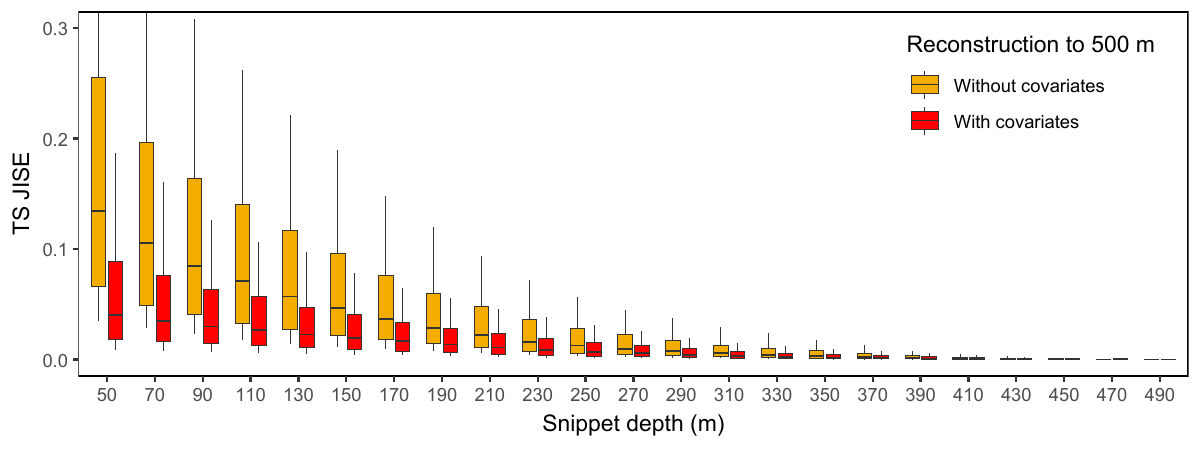}
	\caption{Joint integrated squared errors ($\mathrm{JISE}_m$) for temperature-salinity reconstruction down to 500 m, for various values of snippet depth $z_{max}$ with or without spatial covariates. Boxplot limits correspond to the 0.1, 0.25, 0.5, 0.75, and 0.9 quantiles.}
	\label{FIG:MSE_yn}
\end{figure}

\autoref{FIG:MSE_yn}a summarises the temperature–salinity $\mathrm{JISE}_m$ evaluated for a snippet depth $z_{max} \in \{50, \dots, 490\}$ m, with 20 m increments, and whether geographical covariates are used in the prediction process. In both cases, reconstruction errors decrease as depth $z_{max}$ increases and approaches the reconstruction depth of 500 m. Incorporating geographical location as a covariate in the reconstruction step significantly reduces the JISE$_m$, although the difference in error decreases as snippet depths approach 500 m, where errors become very small. 
Taken separately, when the $M$ profiles are truncated at a snippet depth $z_{\max} = 250$ m, the temperature reconstruction errors yield to a mean value (sd value) of 
$\mathrm{IE}_m^T$ of 0.23 $^\circ\mathrm{C}$ (0.17) which decreases to 0.16 $^\circ\mathrm{C}$ (0.12) when location is included as a covariate. For salinity, decrease in $\mathrm{IE}_m^S$ is also observed moving in average from 0.03 psu (0.02) to 0.02 psu (0.02) with covariate.

\begin{figure}[H]
	\centering
	\includegraphics[width=1\linewidth]{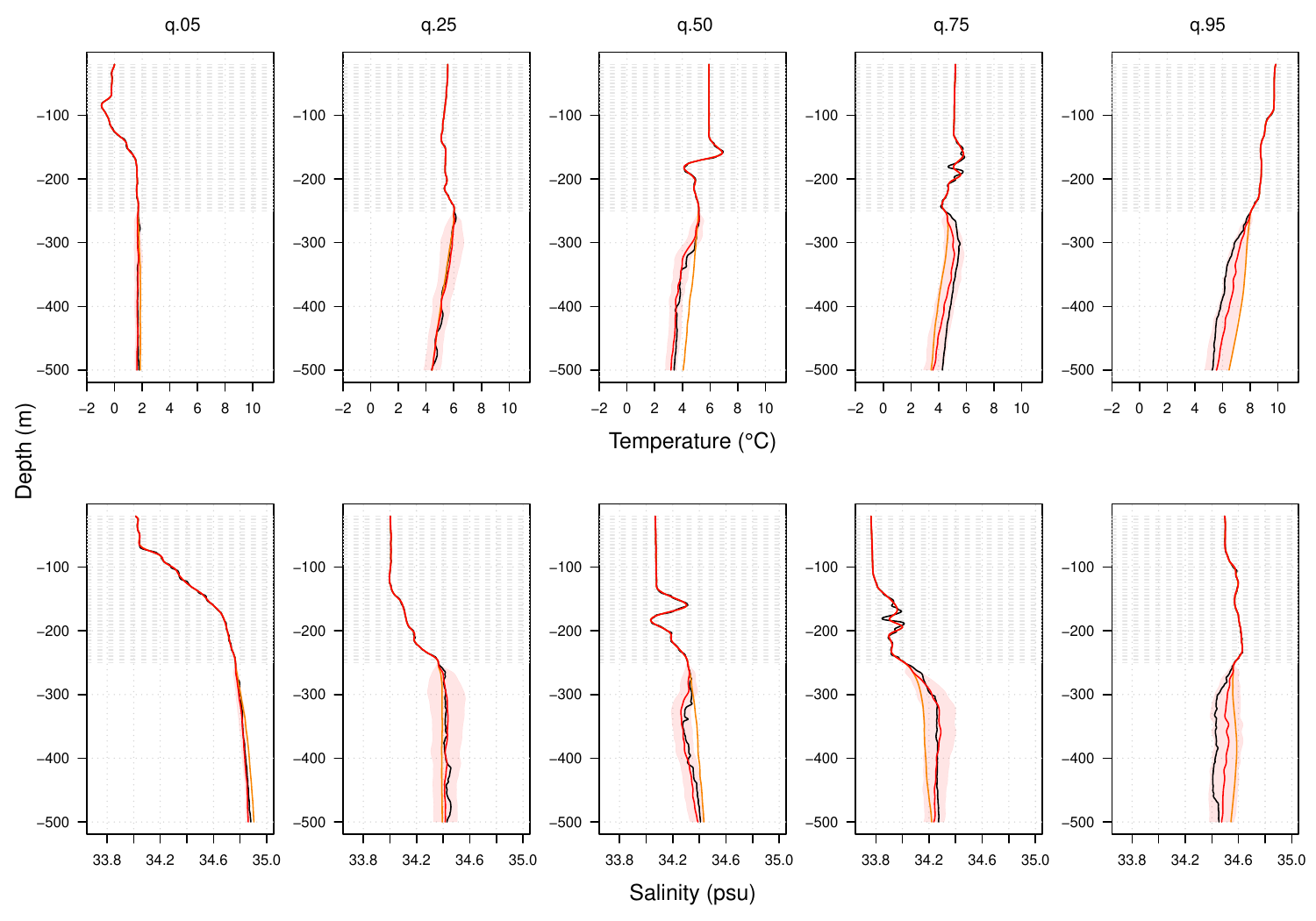}
	\caption{Simultaneous predictions of temperature and salinity profiles using $Q = 100$ components. Black curves display observed profiles. Red curves with 95 \% pointwise confidence intervals are reconstructed profiles when including spatial covariate (orange curves: no covariate). Horizontal dashed lines indicate the position of raw data used for prediction. From left to right, pairwise curves are sorted by the increasing temperature-salinity JISE$_m$ quantiles of order q = 0.05, 0.25, 0.50, 0.75, and 0.95.}
	\label{FIG:WB100}
\end{figure}

\begin{figure}[H]
	\centering
	\includegraphics[width=.8\linewidth]{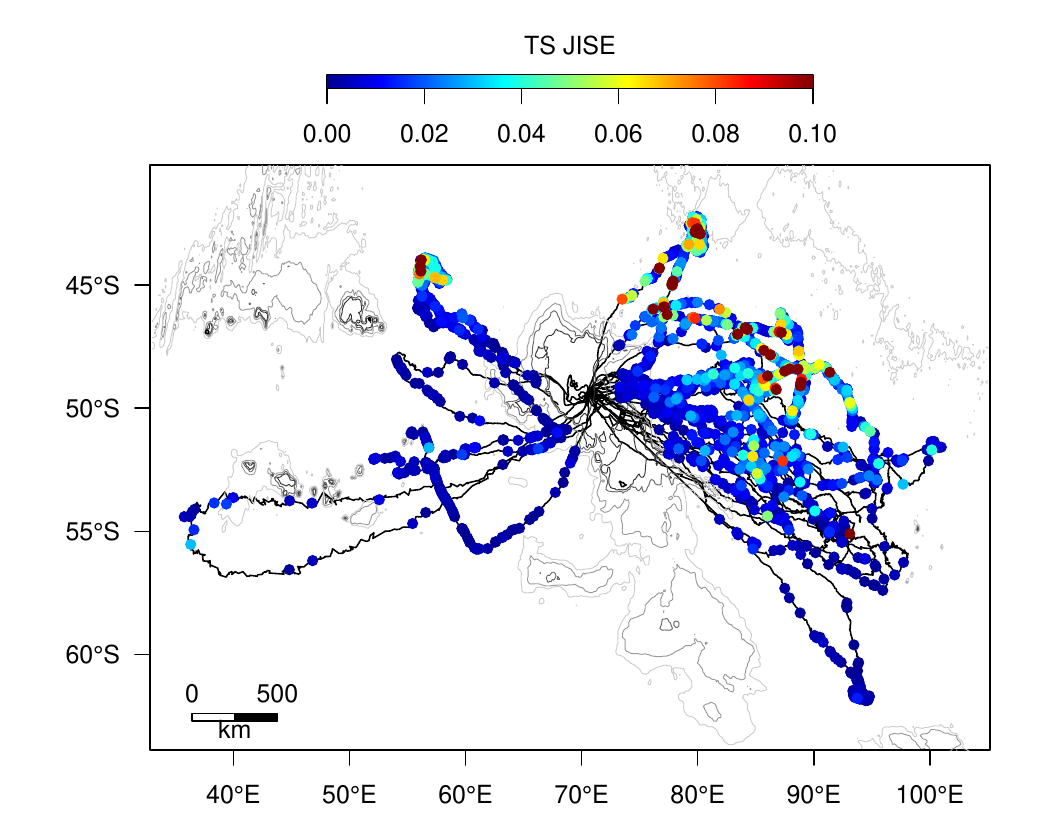}
	\caption{Elephant seal trajectories coloured by the magnitude of the TS joint integrated squared error (JISE$_m$). Snippet profiles are truncated at 250 m and reconstructed down to 500 m, with fPCA performed conditionally on location. JISE$_m$ values are not directly comparable with degrees Celsius ($^\circ$C) or practical salinity units (psu).}
	\label{FIG:MSE}
\end{figure}

Bivariate snippet profiles truncated at depth $z_{max} = 250$ m and reconstructed down to 500 m (conditionally to the location) are shown in \autoref{FIG:WB100} for $Q = 2K = 100$. The segment between the lower bound (20 m) and the snippet depth (i.e., the observed portion used for reconstruction) is accurately predicted, with the reconstructed curves closely overlapping the observed profiles. The reconstruction accuracy over the interval $[z_{max}, 500]$ m generally matches the observed profiles, but seem to depend on the profile shape. The presence of vertical shear, associated with the intrusion of distinct water masses at specific depths, appears to explain local errors when reconstructing profiles. Consistently, higher JISE$_m$ values are observed in the northern and eastern parts of the region visited by the elephant seals (\autoref{FIG:MSE}b), where strong currents and mesoscale activity increase variability in profile structure \citep{park2008large,kim2014variability}.

\begin{figure}[H]
	\centering
	\includegraphics[width=.95\linewidth]{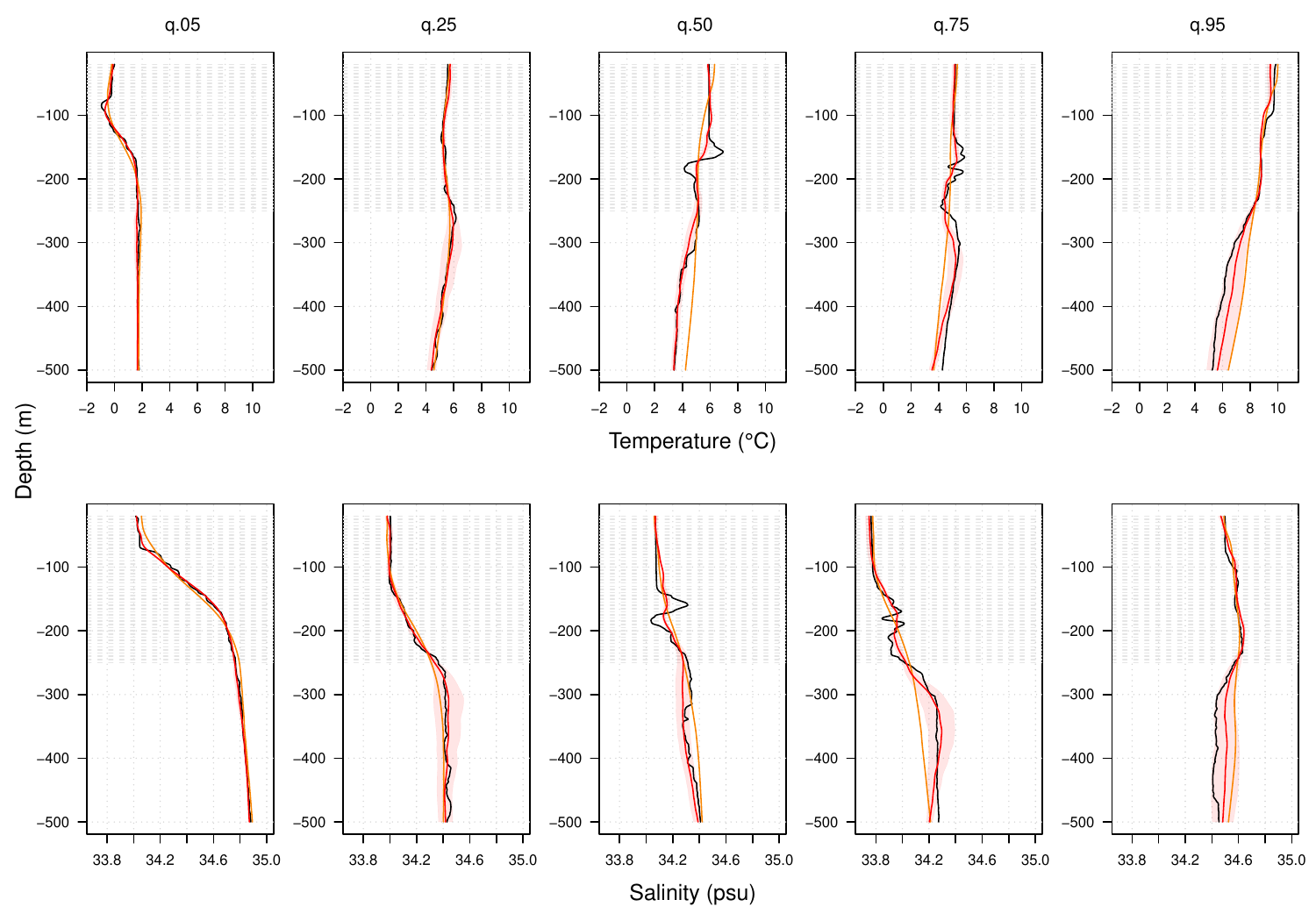}
	\caption{Simultaneous predictions of temperature and salinity profiles using $Q = 5$ components. Black curves display observed profiles. Red curves with 95 \% pointwise confidence intervals are reconstructed profiles when including spatial covariate (orange curves: no covariate). Horizontal dashed lines indicate the position of raw data used for prediction. From left to right, pairwise curves are sorted by the increasing temperature-salinity JISE$_m$ quantiles computed with $Q = 100$ (see comparison with \autoref{FIG:WB100}).}
	\label{FIG:WB5}
\end{figure}

For the five conditional fPCAs performed to reconstruct the profiles in \autoref{FIG:WB100}, the first five principal components explain between 89.54 \% and 97.89 \% of the total variance. The explained variance increases to 99.8 - 99.99 \% when considering the first 50 principal components. The number of retained components $Q$ controls the trade-off between smoothness and fidelity: smaller $Q$ yields a smoother but less flexible fit (underfitting), whereas larger $Q$ captures fine-scale thermohaline structures (see \autoref{FIG:WB5} for comparison with $Q = 5$).
 
\subsection{Profiles reconstruction along an elephant seal track}

\begin{figure}[H]
	\centering
	\includegraphics[width=1\linewidth]{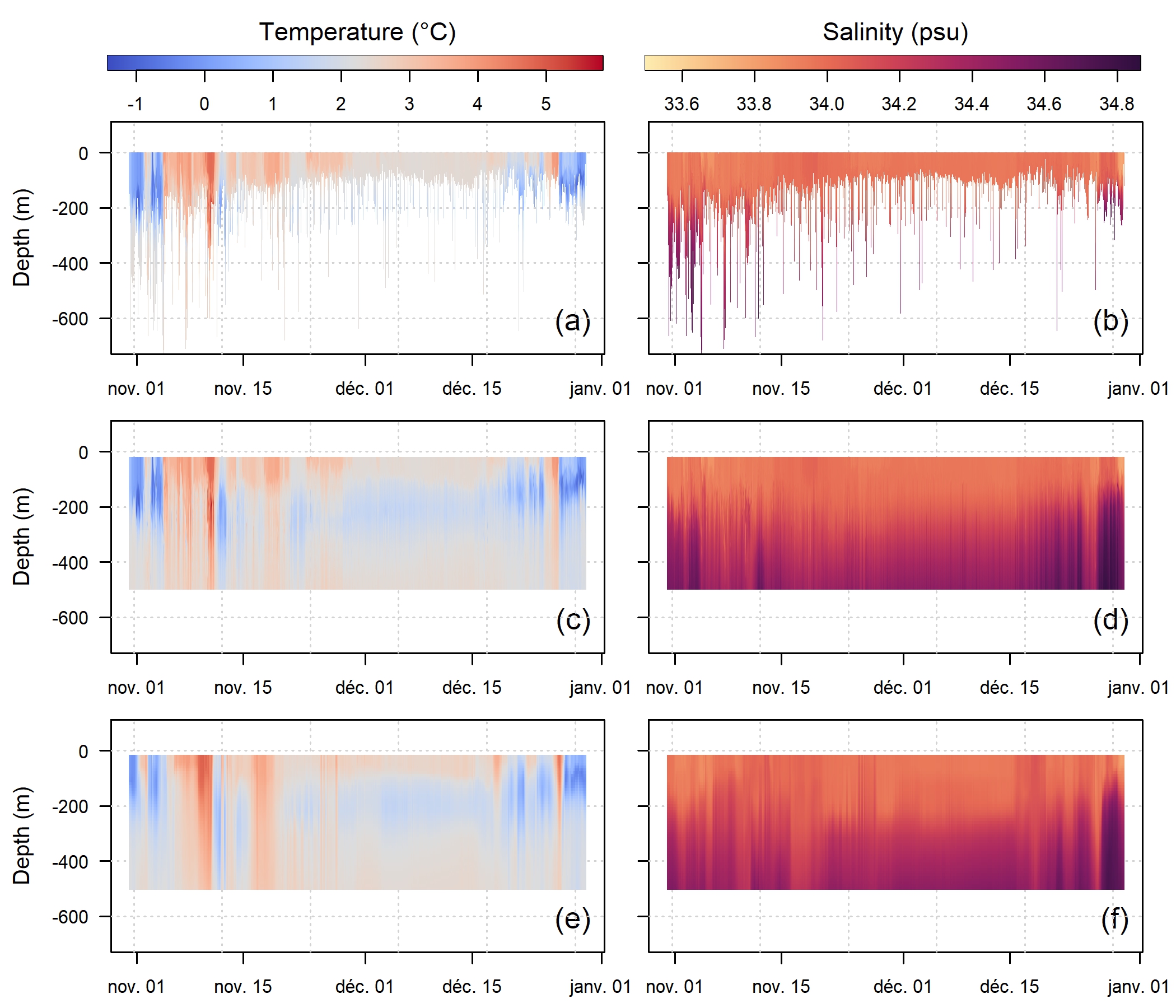}
	\caption{\textbf{(a,b)} In situ temperature (left) and salinity (right) profiles recorded along an elephant seal trajectory, showing variable dive depths. \textbf{(c,d)} Profiles reconstructed down to 500 m using a set of 15,203 profiles collected by 24 females over three years. \textbf{(e,f)} Corresponding temperature and salinity profiles extracted from a reanalysis model at each dive location.}
	\label{FIG:RealReco}
\end{figure}

This section presents the reconstruction method applied to the dataset described in Section \ref{SEC:Data}, which consists of 111,289 temperature and salinity profiles collected along 24 trajectories of female southern elephant seals. The profiles are not deliberately degraded. The maximum depth reached by each profile depends on the dive depth, which varies over time (\autoref{FIG:RealReco}a, b).
Profiles are reconstructed down to 500 m, following the steps highlighted by the dashed box in the workflow shown in \autoref{FIG:WF}.
Profiles to be reconstructed, i.e. those that do not reach 500 m, account for $M =$ 96,086, representing 86.34 \% of the dataset. The remaining $N =$ 15,203 profiles, including profiles with missing geographic coordinates, are used to construct the multivariate fPCAs. These profiles serve as the \textit{learning sample} previously described.

When a pair of TS profiles requiring reconstruction has associated longitude and latitude coordinates, the estimation of the principal component scores is performed conditionally on its location. Among the $M =$ 96,086 profiles to be reconstructed, only 20.51 \% lack geographical coordinates and are thus reconstructed without conditioning on these covariates.
The fPCAs are computed by decomposing the TS profiles using a basis of $K = 50$ penalised B-spline functions defined over the depth range $\mathcal{Z}_{500} = [20, 500]$ m. As described in the previous section, the bandwidth of the Gaussian kernel used in the conditional fPCAs is set to 100 km, corresponding to the typical size of mesoscale oceanic structures.

Profile reconstruction can be performed along each trajectory (e.g., \autoref{FIG:RealReco}c, d), and may be followed by additional post-processing steps. For example, profiles reconstructed down to 500 m from snippets with a maximum depth of only 100 m and lacking geographical information are more prone to reconstruction error (see \autoref{FIG:MSE_yn}).
Such profiles can be excluded from the analysis; in this dataset, they represent 4.72 \% of the profiles to be reconstructed.
In addition, salinity profiles can be adjusted a posteriori to remove density inversions \citep{barker2017stabilizing,siegelman2019correction}.

TS profiles from the GLORYS12V1 model outputs can be used to validate the reconstructed profiles along the seal trajectories. The GLORYS12V1 model (\url{https://doi.org/10.48670/moi-00021}) is a global, eddy-resolving physical ocean and sea-ice reanalysis based on the NEMO model, combining a numerical ocean circulation model with data assimilation of satellite and in situ observations to provide a consistent historical reconstruction of the global ocean state. Model profiles are provided daily at a horizontal resolution of 1/12° and extracted for each dive location (\autoref{FIG:RealReco}e, f). Differences between in situ profiles and co-located model data likely arise from both uncertainties in the model and observational data and from the fact that fine-scale oceanic structures are not always accurately resolved in reanalysis products. The squared distance between in situ profiles recorded by elephant seals and concomitant model profiles is evaluated using JISE$_m$ (equation (\ref{EQ:dist})) over the interval $[20,500]$. JISE$_m$ is significantly smaller between the reconstructed snippet profiles and the model than between the fully observed profiles and the model. Since reconstructed profiles tend to be smoother than raw data, this may partly explain the smaller discrepancies with model data, thus lending further support to the validity of the reconstruction approach.

\section{Discussion and perspectives}

Functional Data Analysis provides a statistical framework particularly well suited for studying data indexed by time or space, which are ubiquitous in the oceanographic field. Within this framework, functional Principal Component Analysis (fPCA) approaches are often used to resolve imputation issues inherent to oceanic and climate observations (e.g., Data INterpolating Empirical Orthogonal Function, \citealp{beckers2003eof}). Missing data may arise from instrumental malfunctions or inaccessible sampling conditions, and can limit the accurate characterisation and understanding of environmental variability \citep{alejo2025missing}. Incomplete data represent a major challenge in satellite remote sensing, where large spatial and temporal gaps are often caused by cloud coverage. While recent studies have investigated spatiotemporal interpolation techniques (e.g., \citealp{palummo2024functional}), much of the existing literature has focused on reconstructing missing values along the temporal dimension (e.g., \citealp{alvera2005reconstruction,shropshire2016storm,ghafarian2018gap,hilborn2018applications,zhou2021research}). In a similar context, \citet{doriot2025distribution} recently applied the Principal Components Analysis through Conditional Expectation (PACE) framework to fill gaps in mixed-layer depth time series between 2007 and 2023 around the Kerguelen Plateau.

Introduced by \citet{yao2005functional}, the PACE framework estimates the mean and covariance functions of sparsely observed data using non-parametric smoothing. The eigenfunctions and eigenvalues of the resulting covariance surface are then obtained through spectral decomposition, and principal component scores are inferred through conditional expectation rather than numerical integration. 
Here, we have adapted the method to densely sampled multivariate snippet functions. In our case, a parametric estimation of the mean and covariance functions is achieved from a pool of $N$ curves observed over the domain of interest $\mathcal{Z} = [z_1,z_L]$ and decomposed into a B-spline basis. We assume that $N$ is sufficiently large to ensure an accurate estimation of the mean and covariance functions, and hence the eigenelements of the spectral decomposition of the covariance function. Under the assumption that the snippet functions belong to the same statistical family as this pool, their principal component scores can be reasonably approximated as the expected values conditional on the observed measurements following equation (\ref{EQ:Yao2D}).

With profiles collected by deep diving mammals, each snippet curve $X_m$ is regularly observed over the subdomain $[z_1, z_{max}] \subset \mathcal{Z}$, with $z_{max} < z_L$, while the remaining portion $(z_{max}, z_L]$ is missing. At each dive, the prey distribution controls the random dive depth $z_{max}$. As underlined by \citet{liebl2019partially}, this configuration does not satisfy the \textit{Missing Completely At Random} assumption. However, unlike most approaches that handle snippet or sparse data, we do not estimate the mean and covariance functions directly from the partially-observed functions (e.g., \citealp{james2000principal,li2010uniform}). Instead, we simplify the problem by selecting the pool of $N$ profiles fully observed over the domain of interest $\mathcal{Z}$, and treat the snippet functions as partially-observed extra curves. Since the mean and covariance functions are estimated from the $N$ profiles in a parametric way, we avoid the common issue of missing information in the far off-diagonal regions of the covariance structure, an issue that has received considerable attention in the literature
(e.g., \citealp{fan2007analysis,cai2016minimax,descary2019recovering,delaigle2021estimating,lin2021basis,lin2022mean}). 
This strategy allows for a more suitable reconstruction of functional trajectories, whereas techniques based on smoothing the empirical covariance function estimated from raw data as performed in PACE tend to produce overly smoothed results and are not suited for handling large datasets.

Our reconstruction approach is only valid if the underlying process governing the observed trajectories of the functional variables is similar across all observations. Because the maximum dive depth $z_{max}$ depends on prey distribution, which in turn is shaped by environmental conditions, sampling bias and data representativeness must be considered. For elephant seals, dives extending down to 500 m occur throughout the entire study region and thus encompass the full range of oceanographic conditions. Although the vertical structure of the water column influences their foraging behaviour, elephant seals also perform drift dives, during which they rest and passively sink to great depths. Drift dives are even more frequent in intense foraging zones where foraging dives can be shallower \citep{crocker1997drift}. These drift dives therefore constitute a unique opportunity to sample the water column more deeply and to complete the partial sampling.

To confirm that the $N$ profiles reaching 500 m and the $M$ snippet profiles belong to the same population, we performed functional linear discriminant analyses between the two groups \citep{li2012dd,cuesta2017dd}, with profiles truncated at depths $z_{max} \in \{50, \dots, 490\}$ m, with 20 m increments, and subsampled to ensure $N = M = 10,000$. For both temperature and salinity, and across all tested snippet depths, between 51 \% and 66 \% of the profiles are correctly classified, indicating weak separability and substantial overlap between the two groups.
 
One way to further develop this aspect would be to explore parametric frameworks designed for functional data displaying non-Gaussian characteristics that vary across space or time (e.g., \citealp{staicu2012modeling}). Several studies have also focused on nonparametric approaches to account for correlation structures in functional data (e.g., \citealp{li2007nonparametric,paul2011principal}). The PACE framework has, for instance, been extended to spatial PACE (SPACE), reconstructing individual curves by explicitly estimating the spatial correlation among functional observations \citep{liu2017functional}. Alternative reconstruction strategies could also be considered, such as spatialised functional models (e.g., \citealp{nerini2010cokriging,menafoglio2016kriging,nerini2022extending}).
Here, we account for the spatial dependence between observations by including geographical location as a covariate in the reconstruction process, following the work of \citet{cardot2007conditional}. By adding known factors that explain part of the variability among curves, we significantly reduce the temperature-salinity joint integrated squared error and account for the three dimensions of our spatial observations (horizontal and vertical).

In the simulation study, we observed that reconstruction accuracy depends on (i) the number of retained components $Q$, (ii) the snippet maximum depth $z_{max}$, (iii) the geographic location of the profile, and (iv) the profile shape (e.g., shear layers). 
While not perfect, the proposed reconstruction framework offers a practical way to mitigate the irregular sampling inherent in marine-mammal observations. It offers a fast, elegant, and straightforward approach to reconstruct bivariate functions simultaneously, by accounting for the covariance between variables. Beyond the snippet-function case, our approach could also be applied to sparse functional data, as originally developed by \citet{yao2005functional}. Although we illustrate it with bivariate hydrological profiles, the method extends to a wide range of functional data types, including time-indexed series \citep{doriot2025distribution}, and to various multivariate functions. It can also accommodate functional variables ranging on different domains (e.g., active acoustic profiles, \citealp{izard2024decomposing}). Finally, when both variables are available in the learning sample, cross-variable prediction is feasible: an entire function of one variable could be inferred from the other (either full or snippet) thereby offering a valuable opportunity to complete historical hydrological datasets where only the temperature was often sampled.  

\section{Conclusion}

Instrumented marine mammals, such as elephant seals, provide unique access to the ocean interior and greatly contribute to the understanding of ocean biophysical dynamics. The wide range of head- and back-mounted tags deployed over the last decades has created a growing need for statistical tools capable of extracting information from large multivariate datasets indexed in time and across the three spatial dimensions. 
We have presented a fast method to reconstruct multivariate snippet profiles when partially sampled by deep-diving predators. The simulation study demonstrates how reconstruction performance varies with snippet depth and highlights the benefit of conditioning a functional principal component analysis when incorporating additional covariates. By simultaneously reconstructing 96,086 temperature and salinity profiles down to 500 m using only 15,203 complete profiles, our approach enhances the usability of animal-borne datasets. This framework could also be extended to other data types, such as satellite observations, to address the challenge of missing data and to open new perspectives for studying oceanographic biophysical systems and their fine-scale dynamics.

\subsection*{Acknowledgements}

This work was conducted as part of N. Fonvieille's PhD, funded by the French Ministry of Education and Research. Elephant seal data were collected within the framework of the “Système National d’Observation : Mammifères Échantillonneurs du Milieu Océanique” (SNO-MEMO, PI: C. Guinet). Fieldwork in the Kerguelen Islands was supported by the French Polar Institute (Institut Polaire Français Paul-Émile Victor) under the CyclEleph program (no. 1201, PI: C. Gilbert). We sincerely thank everyone who contributed to the fieldwork in the Kerguelen Islands.

\subsection*{Data Availability Statement}

The datasets and Gitlab repository access for the codes used in this study are available upon request.

\printbibliography

\section*{Supplementary materials}
\beginsupplement

\end{document}